\documentclass[aps,rmp,onecolumn,amsmath,amssymb,floatfix]{revtex4-2}
\usepackage{graphicx}
\usepackage{bm}
\usepackage{hyperref}

\usepackage{braket,bbold,color}
\usepackage[letterpaper,margin=1.5in]{geometry}

\begin{document}

\title{Tutorial: Dirac Equation Perspective\\ on Higher-Order Topological Insulators}

\author{Frank Schindler}
\affiliation{Princeton Center for Theoretical Science, Princeton University, Princeton, 08544, New Jersey, USA}

\begin{abstract}
In this tutorial, we pedagogically review recent developments in the field of non-interacting fermionic phases of matter, focussing on the low energy description of higher-order topological insulators in terms of the Dirac equation. Our aim is to give a mostly self-contained treatment. After introducing the Dirac approximation of topological crystalline band structures, we use it to derive the anomalous end and corner states of first- and higher-order topological insulators in one and two spatial dimensions. In particular, we recast the classical derivation of domain wall bound states of the Su-Schrieffer-Heeger (SSH) chain in terms of crystalline symmetry. The edge of a two-dimensional higher-order topological insulators can then be viewed as a single crystalline symmetry-protected SSH chain, whose domain wall bound states become the corner states. We never explicitly solve for the full symmetric boundary of the two-dimensional system, but instead argue by adiabatic continuity. Our approach captures all salient features of higher-order topology while remaining analytically tractable.
\end{abstract}

\maketitle

\tableofcontents

\section{Introduction}
We begin by introducing some basic notions of condensed matter physics, topological phases, and topological crystalline insulators. We illustrate symmetry-protected topological phases with a toy model in zero spatial dimensions. Various aspects of the material in this tutorial are also covered in Refs.~\onlinecite{fradkin_2013, AsbothIntro, BernevigBook, BernevigNeupertLectures, Neupert2018, Po_2020}.

Units are chosen so that the speed of light and Planck's constant are dimensionless $c = \hbar = 1$.

\subsection{Basic notions}
Whereas particle physics aims to decompose the macroscopic material world into its elementary constituents, the goal of condensed matter physics is to understand how the properties of materials emerge from their microscopic degrees of freedom. In principle, this question can be posed not only for materials, but for any many-body system that is comprised of a macroscopic number of particles. However, condensed phases of matter are particularly interesting, because they are not \emph{adiabatically} (that is, not without a phase transition) connected to the featureless phase at infinite temperature.
This characterization implies that their behavior cannot be simply understood as the sum of their parts~\cite{Anderson393}: for instance, the elementary particles that make out a condensed matter system often reorganize into collective excitations that behave like particles themselves, the so-called quasi-particles, and yet have markedly different properties, such as a different mass or even different exchange statistics~\cite{wilczek1982quantum}. One of the basic insights of condensed matter physics is that there is a multitude of condensed phases beyond the rather coarse classification of matter into solids, liquids and gases, with particularly interesting cases at very low temperatures where quantum mechanical effects come into play.

A crystal is an example of a condensed phase that consists of atomic nuclei and electrons, which for our purposes are its elementary constituents. We assume that the atomic nuclei (often just called atoms) form a crystalline lattice in $D$ spatial dimensions, and move so slowly in comparison to the electrons -- because of their much larger mass -- that we can essentially treat them as a fixed background. Since the atomic lattice breaks spatial rotation and translation symmetry, the crystal realizes a condensed phase that is not adiabatically connected to the infinite-temperature phase (which is perfectly symmetric). From now on, we only focus on the electronic part, which at low temperatures is governed by quantum mechanics\footnote{The formation of the crystalline lattice, on the other hand, can be understood in classical physics via potential energy minimization.}, and determines whether the crystal is an insulator or a conductor, among many other measurable properties.

Since electrons are fermionic particles obeying the Pauli exclusion principle, a conglomerate of electrons such as the one populating a crystal cannot be understood as the sum of its parts even when there are no further interactions: in contrast to bosons, electrons cannot all occupy the same single-particle eigenstate. This property already enables a great number of interesting electronic phases of matter, and so we will completely neglect further interactions between the electrons such as the Coulomb interaction. It turns out that this approximation is often surprisingly good and applies to many real materials.

Quantum properties dominate over thermal fluctuations at low temperatures, and therefore we will consider systems at zero temperature for simplicity, allowing us to focus on ground state properties only. $N$ non-interacting electrons then fill up the $N$-lowest energy eigenstates of the single-particle Hamiltonian due to the Pauli exclusion principle.
To be precise, let $c^\dagger_\alpha$ be the creation operator of an electron in the single-particle eigenstate labelled by $\alpha$. Fermionic exchange statistics are implemented by the anti-commutation relations
\begin{equation}
\{c^\dagger_\alpha, c^{\vphantom{\dagger}}_\beta\} = 2 \delta_{\alpha \beta}, \quad \{c^\dagger_\alpha, c^\dagger_\beta\} = \{c^{\vphantom{\dagger}}_\alpha, c^{\vphantom{\dagger}}_\beta\} = 0.
\end{equation}
We can write a generic non-interacting many-body Hamiltonian that preserves electronic charge as
\begin{equation} \label{eq: manybodyhamiltonian}
H = \sum_{\alpha} (\epsilon_\alpha-\mu) c^\dagger_\alpha c^{\vphantom{\dagger}}_\alpha,
\end{equation}
where $\epsilon_\alpha$ are the eigenvalues of the single-particle Hamiltonian $\mathcal{H}$, and $\mu$ is the chemical potential. The $N$-particle ground state is then given by 
\begin{equation} \label{eq: slaterdetgroundstate}
\ket{\Omega} = \frac{1}{\mathcal{N}} \prod_{\epsilon_\alpha < \mu} c^\dagger_\alpha \ket{0},
\end{equation}
where $\mathcal{N}$ is a normalization factor, $\ket{0}$ denotes the electronic vacuum, and each $\alpha$ appears exactly once as long as $\epsilon_\alpha < \mu$. $\ket{\Omega}$ is called a Slater-determinant state, because the fermionic anti-commutation relations imply that its coefficients in a given basis are given by a determinant.\footnote{In contrast, non-interacting bosons created by $b^\dagger_\alpha$ condense into a Bose-Einstein condensate, where they all populate the same single-particle wavefunction: $\ket{\Omega} \propto (b^\dagger_{\tilde{\alpha}})^N \ket{0}$, where $\tilde{\alpha}$ labels the lowest-energy state of the single-particle Hamiltonian.}

The total number of single-particle states scales linearly with the system volume $V$. An insulator is defined by an energy gap between the ground state and the lowest excited state that survives in the thermodynamic limit, where we take $N \rightarrow \infty$, $V \rightarrow \infty$ while keeping the particle density $N/V$ constant. The condition for a gap in the many-body spectrum of $H$ translates to the absence of single-particle eigenstates with energies at the chemical potential $\mu$. A gap implies that a small external electric field\footnote{By a small electric field, we mean that its energy per particle is much smaller than the gap.} cannot lift electrons to excited single-particle states with nonzero net momentum, and therefore cannot create a current. In contrast, a conductor (a metal) is characterized by the absence of a gap. A (second-order) phase transition, which features a diverging correlation length and a discontinuous change in some ground state properties, implies a gap closing.

\subsection{Topological phases}
\label{sec: introduction}

A condensed matter system realizes a gapped topological phase when its ground state cannot be continuously deformed to a trivial reference state adiabatically, that is, without closing the energy gap to the first excited state. In the context of a crystal, a trivial insulator can be defined as an \emph{atomic limit} where all electrons are tightly bound to the atoms, that is, as a state that is adiabatically connected to a Slater product over single-electron wavefunctions whose weight is exponentially localized around the atomic sites in position space~\cite{Bradlyn17}. For this definition, it is important to assume periodic boundary conditions, so that boundary effects do not play a role and only bulk crystal properties are probed. We refer to the gap of a crystal with periodic boundary conditions as the \emph{bulk gap}, and adiabatic manipulations as those preserving it. There may be multiple trivial phases: for instance, in a crystal containing multiple types of atoms (such as rock salt, $\mathrm{NaCl}$), the states formed by electrons localized around each atom are equally valid atomic limits (in real materials, however, at most one atomic limit is energetically favored).

\begin{figure}[t]
\centering
\includegraphics[width=0.55\textwidth,page=1]{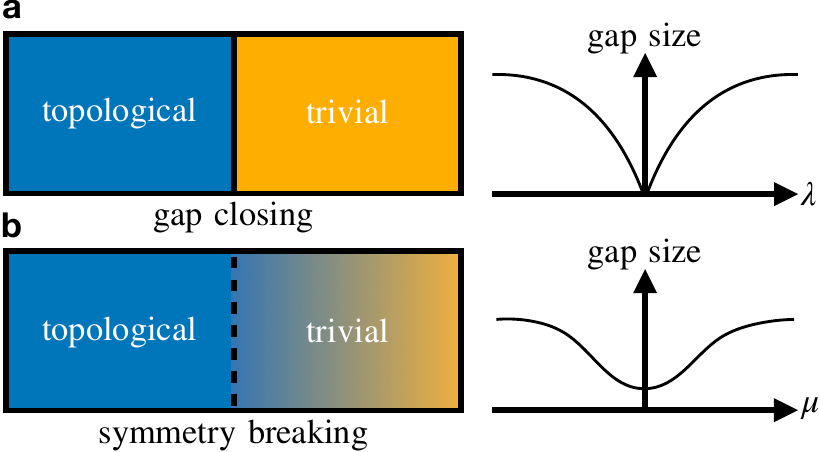}
\caption{Schematic phase diagrams of a symmetry-protected topological phase. (a)~When all symmetries are maintained, interpolating from the topological to the trivial parameter regime necessitates a gap closing and thereby a phase transition. The right panel shows how the (absolute value of the) gap evolves with a tuning parameter $\lambda$, where the Hamiltonian $H(\lambda)$ is symmetric for all choices of $\lambda$. (b)~Upon symmetry breaking, a smooth (adiabatic) transition without gap closing becomes possible. The right panel shows how the gap evolves with a tuning parameter $\mu$, where $H(\mu)$ is symmetric for $\mu<0$ but the symmetry is broken for $\mu > 0$.
}
\label{fig: schematic_spt}
\end{figure}

The concept of topological phases can be refined by demanding that symmetry preservation is necessary to prevent an adiabatic path to a trivial phase, one then speaks of a \emph{symmetry-protected topological phase} (a pedagogical review is provided by Ref.~\onlinecite{senthil15}). (See Fig.~\ref{fig: schematic_spt}.) For example, it may be possible to adiabatically deform the ground state of all electrons in a crystal to an atomic limit that breaks the crystalline inversion symmetry (which maps a coordinate $x$ in the crystal lattice to $-x$), but to no other atomic limit. The state then realizes a \emph{topological crystalline insulator}~\cite{Fu11} protected by inversion symmetry. The adjective ``crystalline" here just means that a crystalline symmetry, in addition to the bulk gap, is necessary to stabilize the topological phase. When the symmetry is not crystalline, such as time-reversal, the convention is to simply call the system a topological insulator. Historically, topological insulators protected by time-reversal symmetry came before topological crystalline insulators, here we nevertheless focus on the latter class of systems because they are conceptually much simpler. There is only one known example of a non-interacting topological phase that requires no symmetries at all for its protection: the Chern insulator, or integer quantum Hall effect~\cite{Thouless82}.

To illustrate these concepts, consider a zero-dimensional toy model of a spinless electron occupying one of two orbitals in an atom, an $s$ and a $p$ orbital, with single-particle Hamiltonian
\begin{equation} \label{eq: 0DBlochExample}
\mathcal{H} = \alpha \sigma_x + \beta \sigma_y + \gamma \sigma_z, \quad \epsilon_\alpha = \pm \sqrt{\alpha^2+\beta^2+\gamma^2},
\end{equation}
where $\sigma_a$, $a=x,y,z$, are the $2 \times 2$ Pauli matrices, and there are only two energy levels labelled by  their sign $\alpha = \pm$. Let inversion symmetry be represented by $\mathcal{I} = \sigma_z$, so that the $s$ orbital corresponds to the state $\ket{s} = (1,0)^\mathrm{T}$, while the $p$ orbital corresponds to $\ket{p} = (0,1)^\mathrm{T}$ ($\mathrm{T}$ takes the transpose). For $|\gamma| \gg |\alpha|, |\beta|$, for instance, the spectrum of $\mathcal{H}$ is gapped, and therefore also that of the corresponding many-body Hamiltonian $H$ in Eq.~(\ref{eq: manybodyhamiltonian}) at chemical potential $\mu = 0$. Let us now define $\ket{s}$ as the trivial reference state, it is the ground state of $\mathcal{H}$ for $\gamma = -\infty$.\footnote{Note that in zero dimensions, there is no notion of wavefunction localization or atomic limit, we therefore have to fix the trivial phase by hand. This makes the notion of topological phase somewhat artificial, but allows to demonstrate it in its simplest possible variant.} Then the system realizes an inversion symmetry-protected topological phase for large positive $\gamma$: as long as inversion symmetry is not enforced, it is possible to tune $\gamma$ from positive to negative values without a gap closing, which happens only when $\alpha^2+\beta^2+\gamma^2 = 0$, as long as $\alpha$ and $\beta$ are kept non-zero. On the other hand hand, enforcing inversion symmetry via the commutation relation $[\mathcal{H},\mathcal{I}] = 0$ implies that $\alpha$ and $\beta$ vanish, which in turn ensures a gap closing along the interpolation $\gamma(t) = 1 - 2t$, $t \in [0,1]$ that exchanges $\ket{p}$ with $\ket{s}$ as the lowest-energy eigenstate of $\mathcal{H}$. We therefore find that there are no topological insulators in zero spatial dimensions that are protected by the bulk gap alone, but that there do exist topological crystalline insulators protected by inversion symmetry.

Since the vacuum can be viewed as a trivial phase\footnote{The vacuum corresponds to an atomic insulator where the atoms are infinitely far apart.}, the boundary of a topological phase generically hosts a phase transition (a phase transition must occur to go between topological and trivial) and thereby a gap closing, as long as it respects the protecting symmetries.\footnote{One might ask why the usual non-topological phases, such as ferromagnets, do not host boundary phase transitions by the same argument. The reason is that these are characterized by local order parameters (such as magnetization) that can smoothly go to zero at the sample boundary, with an energy cost that scales sub-extensively with the system size. Topological insulators, on the other hand, are characterized by non-local and quantized topological invariants that cannot change continuously.} The most experimentally relevant kinds of topological insulators are therefore characterized by an insulating (gapped) bulk and a conducting (gapless) boundary. If the protecting symmetry is not crystalline, all boundaries preserve it by default and are therefore gapless, implying that the $D$-dimensional insulating bulk is surrounded by a $(D-1)$ dimensional conducting surface -- this is the celebrated \emph{bulk-boundary correspondence}. Topological crystalline insulators, on the other hand, generically have surfaces that do not preserve the relevant crystalline symmetry on their own, and can therefore be gapped. The crystalline symmetry may still enforce the presence of $(D-k)$-dimensional gapless states with $k > 1$, thereby giving rise to \emph{higher-order topological insulators}, where $k$ is referred to as the order.

In this tutorial, our aim is to give a pedagogical introduction to first- and higher-order topological crystalline insulators from the point of view of their effective low-energy description. Our main tool will be the Dirac equation, which describes elementary electrons in particle physics. In a particularly tractable realization of the principle of emergence discussed at the beginning of this introduction, we will see how variants of the Dirac Hamiltonian arise in the low-energy description of topological (crystalline) insulators (see also chapter V.5 of Ref.~\onlinecite{zee2010quantum}). We will then explain how they can be used to argue for the stability of (higher-order) topological phases, and in particular for the presence of protected boundary states.

\section{Dirac approximation of topological crystalline insulators} \label{sec: diracapproxintro}
In this section, we introduce the Dirac equation in the context of the tight-binding description of electron movement in crystals. We also fix some basic notation and conventions that will be extensively used later on. For simplicity, we only treat one-dimensional (1D) systems in this section, the generalization to higher dimensions being straightforward (it mostly involves making all position and momentum coordinates vector-valued, and replacing products between them by dot products).

\subsection{Tight-binding description} \label{sec: tb_basics}
Consider a 1D crystal.\footnote{Naively, the Mermin-Wagner theorem prevents the formation of crystals, which break the continuous translational symmetry of space down to the discrete subgroup of lattice translations, in dimensions smaller than three. Nevertheless, one-dimensional crystals exist in nature, either because they are not perfectly one-dimensional (their atoms do not all lie along a single line) such as is the case for polymer chains, or because they are part of three-dimensional structures, such as is the case for adatom chains.} We may model it as an infinitely long chain of repeated atom arrangements, the unit cells. In the tight-binding approximation, each unit cell hosts a finite number of sites that can be occupied by electrons. Denote by $c^\dagger_{x i}$ the operator that creates an electron at site $i$, $i = 1 \dots M$, in the unit cell at position $x$, $x = 1 \dots L$ (we take $x$ to be dimensionless so that $L$ is just the total number of unit cells in the sample). Assuming only nearest-neighbor hoppings between the sites (in a microscopic treatment, these hoppings would be derived from transition matrix elements of the microscopic electronic Hamiltonian between states that are well-localized on the individual sites), the Hamiltonian takes the form
\begin{equation} \label{eq: realspaceHamiltonian}
    H = \sum_{x i j} h^{\vphantom{\dagger}}_{i j} c^\dagger_{x i} c^{\vphantom{\dagger}}_{x j} + t^{\vphantom{\dagger}}_{i j} c^\dagger_{x+1,i} c^{\vphantom{\dagger}}_{x j} + t^{*}_{i j} c^\dagger_{xj} c^{\vphantom{\dagger}}_{x+1, i}.
\end{equation}
(Here and in the following, we will set the chemical potential $\mu$ to zero for simplicity.) By construction, $H$ preserves a global $U(1)$ symmetry that is represented by
\begin{equation}
c^\dagger_{x i} \rightarrow e^{\mathrm{i}\phi} c^\dagger_{x i}, \quad c^{\vphantom{\dagger}}_{x i} \rightarrow e^{-\mathrm{i}\phi} c^{\vphantom{\dagger}}_{x i},
\end{equation}
with $\phi \in (0,2\pi]$ an arbitrary phase. The infinitesimal generator of this symmetry is the total charge,
\begin{equation} \label{eq: total_charge}
Q = \sum_{x i} c^\dagger_{x i} c^{\vphantom{\dagger}}_{x i},
\end{equation}
so that we have $e^{\mathrm{i}\phi Q} c^\dagger_{x i} e^{-\mathrm{i}\phi Q} = e^{\mathrm{i}\phi} c^\dagger_{x i}$. Let $\ket{\Omega}$ be the many-body ground state of $H$, and perform a $U(1)$ transformation by the angle $2\pi$. Evidently, such a transformation must be equal to the identity transformation, and so the ground state, which transforms as $\ket{\Omega} \rightarrow e^{\mathrm{i} 2\pi Q} \ket{\Omega}$, must have integer charge $Q \in \mathbb{Z}$. We will see in Sec.~\ref{sec: filling_anomaly} how the boundary of a topological phase can break this quantization constraint and host fractional charges at the expense of a topologically nontrivial bulk.

Let us for now return to exploring the general properties of the Hamiltonian $H$. Assuming periodic boundary conditions, we can exploit its discrete translational symmetry and partially diagonalize it by introducing the crystal momenta $k = (1 \dots L) \frac{2\pi}{L}$, and the correspondingly Fourier-transformed creation operators
\begin{equation}
    c^\dagger_{k i} = \sum_x e^{-\mathrm{i} k x} c^\dagger_{x i}, \quad c^\dagger_{x i} = \frac{1}{L} \sum_{k} e^{\mathrm{i} k x} c^\dagger_{k i}.
\end{equation}
Later on, we will often implicitly take $L \rightarrow \infty$, in which case all sums over $k$ should be interpreted as integrals over the first Brillouin zone $k \in (0,2\pi]$.
Equation~(\ref{eq: realspaceHamiltonian}) becomes
\begin{equation}
    H = \frac{1}{L} \sum_{k i j} \mathcal{H}_{i j}(k) c^\dagger_{k i} c^{\vphantom{\dagger}}_{k j},
\end{equation}
where we introduced the Bloch Hamiltonian
\begin{equation} \label{eq: 1DgeneralBloch}
    \mathcal{H}_{i j}(k) = h_{i j} + t_{i j} e^{\mathrm{i} k} + t^{*}_{j i} e^{-\mathrm{i} k}.
\end{equation}
The eigenvalues of $\mathcal{H}(k)$ give the single-particle energy spectrum as a function of crystal momentum $k$, the so-called electronic band structure. That is, in the case of a periodic one-dimensional crystal, the eigenstate label $\alpha$ that appears in Equations~(\ref{eq: manybodyhamiltonian}) and~(\ref{eq: slaterdetgroundstate}) is a composite label $\alpha = (k,n)$ that specifies the crystal momentum $k$ together with the index of the $n$-th eigenstate of $\mathcal{H}(k)$, where $n = 1 \dots M$ (there are $M$ bands). [Equation~(\ref{eq: 0DBlochExample}) can be viewed as a Bloch Hamiltonian in zero dimensions, where there is no momentum $k$, and with $M=2$ sites in the unit cell.]

\subsection{Two-band models and topology in the absence of symmetry}
In the remainder of this section, we consider the simplest nontrivial case, where the unit cell has two atomic sites, $M=2$. We can then expand the Bloch Hamiltonian $\mathcal{H}(k)$ in terms of the identity matrix and the Pauli matrices (something that is possible for any Hermitian $2 \times 2$ matrix):
\begin{equation} \label{eq: generaltwobandH}
    \mathcal{H}(k) = \epsilon_0(k) \mathbb{1} + \bm{d}(k) \cdot \bm{\sigma},
\end{equation}
where we used scalar product notation and introduced the real-valued vector $\bm{d}(k)$ with components $d_a(k)$, $a=x,y,z$. The spectrum of $\mathcal{H}(k)$ can be easily inferred by calculating $[\mathcal{H}(k)-\epsilon_0(k)\mathbb{1}]^2$, which turns out to be proportional to the identity matrix $\mathbb{1}$. It is given by
\begin{equation} \label{eq: generalDispersion}
    \epsilon_\pm(k) = \epsilon_0(k) \pm \sqrt{\bm{d}(k) \cdot \bm{d}(k)}.
\end{equation}
We see that the condition for an energy gap between the two bands $\epsilon_+(k)$ and $\epsilon_-(k)$ translates to the inequality $d_a(k) \neq 0$, $a=x,y,z$. Moreover, we can always adiabatically tune $\epsilon_0(k) \rightarrow 0$ without closing the gap between the two energy levels, and so all two-band insulators that only differ by the choice of $\epsilon_0(k)$ are topologically equivalent. We set $\epsilon_0(k) = 0$ in the following.

\begin{figure*}[t]
\centering
\includegraphics[width=1\textwidth,page=2]{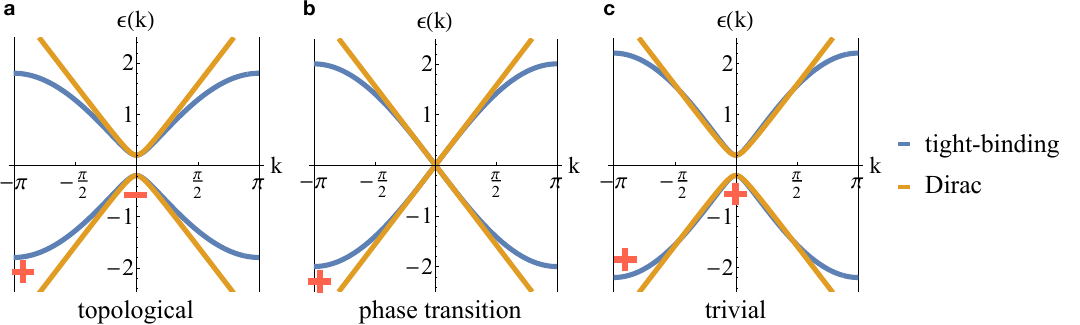}
\caption{Dirac approximation of a 1D inversion-symmetric tight-binding model at its topological phase transition. The original electronic structure [Eq.~\eqref{eq: SSHmodel}] is shown in blue, while the Dirac spectrum [Eq.~\eqref{eq: SSHDiracHamiltonian}] is shown in orange. The signs of the occupied inversion eigenvalues, when well-defined, are indicated in red (at the gap closing, we cannot unambiguously fix the sign). (a)~The topological regime ($m>0$) is characterized by inversion eigenvalues that have opposite sign at $k=0$ and $k=\pi$. (b)~At the phase transition ($m=0$), the inversion eigenvalue at $k=0$ is exchanged. (c)~The trivial regime ($m<0$) is characterized by equal occupied inversion eigenvalues. Importantly, the Dirac approximation reproduces the sign switch at $k = 0$, all other details of the band structure are unimportant from the point of view of crystalline topology.
}
\label{fig: DiracTBcomparison}
\end{figure*}

We now come to topology. Due to the above criterion for an energy gap, the space of all one-dimensional two-band insulators is isomorphic to the space of all vector-valued functions $\bm{d}(k)$ that do not map any momentum $k \in (0,2\pi]$ to the zero vector $\bm{0}$, that is, $\bm{d}(k) \in \mathbb{R}^3 \setminus \{\bm{0}\}$. Since the Brillouin zone has the topology of a circle, or one-sphere $S^1$ (after all, the replacement $k \rightarrow k + 2\pi$ has no effect), a Bloch Hamiltonian is characterized by a collection of vectors $\bm{d}(k)$ that satisfies $\bm{d}(2\pi) = \bm{d}(0)$. It is easy to see that all such loops in $\mathbb{R}^3$ can be contracted to a point $\bm{d}(k) = \bm{d}^*$ without ever touching the origin $\bm{0}$. Comparing with Eq.~(\ref{eq: 1DgeneralBloch}), this implies that any constant $\bm{d}^*$ gives rise to a corresponding atomic limit Hamiltonian $\mathcal{H}(k) = \mathcal{H}^*$ that is independent of $k$ and does not feature any hoppings between unit cells.\footnote{Note that our notion of (trivial) atomic limit, which is characterized by vanishing hopping terms, depends on the choice of unit cell.} We deduce that all one-dimensional two-band insulators are topologically trivial absent symmetries.

\subsection{Crystalline topology by inversion symmetry} \label{subsec: inversion_topology_general}
Here and in the following, we will specify only a single kind of trivial reference phase: the atomic limits with a Bloch Hamiltonian independent of $k$. There may be other atomic limits (remember from Sec.~\ref{sec: introduction} that atomic limits are simply insulators whose ground state can be decomposed into exponentially localized electronic wavefunctions). For instance, the Hamiltonian belonging to an atomic limit where all electrons are exponentially localized around atomic sites at the boundary of the unit cell necessarily has to be $k$-dependent. From our point of view, we will treat such phases as topological as long as they cannot be adiabatically transformed into a trivial reference phase, similar to the more canonical examples of topological phases that do not have a representation in terms of exponentially localized electrons at all.\footnote{In the literature, such additional inequivalent atomic limits are called obstructed atomic limits (OALs), here we gloss over the distinction between topological phases, OALs, and the recently introduced fragile phases~\cite{AshvinFragile} -- these can all be viewed as topological by the same criterion, namely that we cannot adiabatically deform them to a trivial atomic limit that has vanishing hopping terms.}

In order to stabilize nontrivial 1D insulators, we need to incorporate symmetries. As mentioned in Sec.~\ref{sec: introduction}, the simplest example of a crystalline symmetry is given by spatial inversion that takes $x$ to $-x$. When each unit cell contains two atoms that are mapped onto each other by inversion, it is represented by $\mathcal{I} = \sigma_x$. At the same time, inversion symmetry flips a crystal momentum $k$ to $-k$. The condition for inversion symmetry in 1D is therefore that the Bloch Hamiltonian satisfies 
\begin{equation}
\mathcal{I} \mathcal{H}(k) \mathcal{I}^\dagger = \mathcal{H}(-k).
\end{equation}
[This condition is equivalent to the requirement that the many-body inversion operator commutes with the many-body Hamiltonian in Eq.~(\ref{eq: manybodyhamiltonian}).] In our two-band example, this constraint enforces that $d_y (k) = - d_y (-k)$ and $d_z (k) = -d_z (-k)$ are odd functions of $k$, while $d_x (k) = d_x (-k)$ is even. Moreover, we find that the crystal momenta $\bar{k} = 0,\pi$, which satisfy $\bar{k} = - \bar{k} \text{ mod } 2\pi$, are special in that, at these momenta, we have the commutation relation $[\mathcal{H}(\bar{k}),\mathcal{I}] = 0$. 

These constraints are sufficient to protect topological crystalline insulators: Consider the many-body ground state at zero chemical potential. In our example [Eq.~\eqref{eq: generaltwobandH}], this situation corresponds to occupying a single site per unit cell with an electron (half filling), and therefore, a single occupied energy band (each band has $L$ available crystal momenta that may be occupied, and there are $L$ unit cells with two sites each).\footnote{In the present example, half filling implies that the band gap of $\mathcal{H}(k)$ induces an energy gap above the many-body ground state of $H$. It is important to keep in mind that the filling (the number of electrons in the sample), or alternatively the chemical potential, is just as important as the presence of a band gap in order for a system to be insulating.} The occupied band is composed of Bloch states $\ket{u(k)}$ that satisfy 
\begin{equation}
\mathcal{H}(k) \ket{u(k)} = \epsilon_-(k) \ket{u(k)}. 
\end{equation}
At the high-symmetry momenta $\bar{k}$, these states can furthermore be chosen as eigenstates of inversion symmetry: 
\begin{equation}
\mathcal{I} \ket{u(\bar{k})} = \lambda(\bar{k}) \ket{u(\bar{k})}. 
\end{equation}
Due to $\mathcal{I}^2 = 1$ we have $\lambda(\bar{k}) = \pm 1$ -- the inversion eigenvalues are quantized to only two allowed values (this quantization is a key ingredient for topology). Now, an insulating state where the two occupied inversion eigenvalues differ, $\lambda(0) = -\lambda(\pi)$, is clearly not adiabatically connected to any trivial atomic limit: all constant Bloch Hamiltonians $\mathcal{H}(k) = \mathcal{H}^*$ have equal occupied inversion eigenvalues, and these cannot be changed without either closing the gap to the unoccupied band with energy $\epsilon_+(k)$, or breaking inversion symmetry. As an example, consider the inversion-symmetric Bloch Hamiltonian for $[d_x(k),d_y(k),d_z(k)] = [\Delta + \cos k,\sin k,0]$, which reads
\begin{equation} \label{eq: SSHmodel}
    \mathcal{H}_{\text{SSH}}(k) = (\Delta + \cos k) \sigma_x + \sin k \sigma_y.
\end{equation}
This Hamiltonian is in the topological phase for $|\Delta| < 1$, and in the trivial phase otherwise. [Comparing with Eq.~(\ref{eq: generalDispersion}), the energy dispersion reads 
\begin{equation}
\epsilon_\pm(k) = \pm \sqrt{\Delta^2 + 2\Delta \cos k + 1}, 
\end{equation}
and so there is a gap closing at $k=\pi$ for $\Delta = 1$, and one at $k=0$ for $\Delta=-1$.] In fact, $\mathcal{H}_{\text{SSH}}(k)$ is closely related to the famous Su-Schrieffer-Heeger (SSH) model of polyacetylene~\cite{SSH} (for a derivation, see Sec.~\ref{sec: SSH_overarching}). We will show in Sec.~\ref{subsec: ssh_edgestate_derivation} that the topological phase is characterized by the presence of end states in an open chain geometry. The model in Eq.~(\ref{eq: SSHmodel}) provides the archetypical model of topological insulators, from which many more phases can be constructed by dimensional and symmetry enhancement.

\subsection{Derivation of the Dirac Hamiltonian} \label{sec: dirac_ham_ssh_derivation}
We have found that there are 1D topological crystalline insulators protected by inversion symmetry. To explore their physical properties with respect to boundaries, we would like to solve for the spectrum and eigenstates of Eq.~(\ref{eq: SSHmodel}) in geometries with open rather than periodic boundary conditions. This can be done numerically via exact diagonalization, but it is more illuminating to find an approximate analytical solution, in particular in order to determine which properties are specific to the model at hand, and which are general features of the topological phase.\footnote{In fact, an exact analytical solution is possible for the specific model considered here, but it is not particularly illuminating as it only works in this fine-tuned case. We are not interested in the minute details of solutions to one particular model, but rather want to determine the universal topological features of the corresponding phase of matter. For this purpose, the approximation we will make is sufficient, and has the advantage that it can be generalized to arbitrary tight-binding models.} 

In order to make the problem tractable, we first need to determine the salient feature distinguishing the topological phase from the trivial phase. Fixing $\Delta < -1$ for the trivial reference phase (that is, the atomic limit where all electrons are tightly bound in the center of each unit cell as a symmetric, even-parity superposition), this feature is clearly the inversion eigenvalue at $k=0$: In the trivial phase, we have $\lambda(0) = \lambda(\pi) = 1$, while in the topological phase we have $\lambda(0) = -\lambda(\pi) = -1$ [remember that $\lambda(\bar{k})$ is the eigenvalue of inversion symmetry $\mathcal{I}$ in the occupied, lower-energy eigenstate of $\mathcal{H}_{\text{SSH}}(k)$]. To capture the topological phase, it is therefore sufficient to perform a first-order Taylor expansion of $\mathcal{H}_{\text{SSH}}(k)$ around $k=0$, arriving at the Hamiltonian
\begin{equation} \label{eq: SSHDiracHamiltonian}
    \mathcal{H}_{\text{Dirac}}(k) = m \sigma_x + k \sigma_y,
\end{equation}
where we identified $m = \Delta + 1$.
This is the Hamiltonian of a Dirac fermion with mass $m$ that propagates in $1+1$ spacetime dimensions.\footnote{The reader might be more familiar with the real-space Dirac equation in $3+1$ spacetime dimensions, which is constructed using $4 \times 4$ $\gamma$-matrices instead of the $2 \times 2$ Pauli matrices. The formal requirement for a Dirac Hamiltonian is that all matrices that appear anti-commute (form a Clifford algebra) and square to $\mathbb{1}$. This requirement can be satisfied using only $2 \times 2$ matrices in $1+1$ dimensions, but not in $3+1$ dimensions (where there are two more momenta that need to be accomodated).} (See Fig.~\ref{fig: DiracTBcomparison} for a comparison of the tight-binding and Dirac energy band spectrum.) It is somewhat curious that the relativistic Dirac equation appears in the context of an evidently non-relativistic (non Lorentz-invariant) crystal. Nevertheless, the Dirac approximation (also called $\bm{k} \cdot \bm{p}$ approximation) can be studied in its own right for any tight-binding model, and is particularly useful in topological band theory: topological properties are insensitive to the non-universal details of the Bloch Hamiltonian that we discarded by approximating around $k=0$.\footnote{The expansion around $k=0$ is often called long-wavelength or low-energy expansion. Here, it is however purely coincidental that the topological distinction between trivial and topological originates from $k=0$. Had we chosen the atomic limit built from antisymmetric, odd-parity superpositions as our trivial reference phase, the relevant momentum would have been $k=\pi$. This underlines the fact that topological properties do not care about energetics or length scales, they are global properties of the full band structure, and not just of its long-wavelength part.}

Note that by virtue of the Taylor expansion, the momentum variable $k$ loses the Brillouin zone periodicity property it previously enjoyed. It should therefore be properly seen as taking values in $k \in (-\infty, \infty)$ (additionally, we may assume the thermodynamic limit $L \rightarrow \infty$, so that $k$ becomes a continuous variable). However, large values of $k$ are unphysical (the Taylor expansion breaks down), we will therefore only use the predictions of the Dirac approximation in a small window around $k=0$, and deduce the behavior at the remaining momenta by arguments of adiabatic continuity. The advantage that the Dirac approximation has over the original tight binding model is that it effectively ``forgets" the microscopic structure of the crystalline lattice and lives in continuous space\footnote{In this tutorial, we only treat so-called strong topological phases that do not require (discrete) translational symmetry for their protection. Correspondingly, we do not care about the particular lattice structure of the crystal, and may discard it with impunity when going over to the continuum Dirac description.}: in order to go to real space (which we need to study finite sample geometries and boundary effects), we can just make the canonical replacement $k \rightarrow -\mathrm{i} \partial/\partial x$.

In the following, we will use the Dirac equation approach to capture the boundary physics of the one-dimensional SSH model, as well as that of a higher-order topological phase in two dimensions. The general strategy will be the same: 
\begin{enumerate}
\item{Write down the tight-binding model.}
\item{Identify the momentum at which the gap closes between trivial and topological parameter regimes.}
\item{Perform a Taylor expansion around this momentum to first order, thus arriving at a Dirac-type Hamiltonian.}
\item{Use the Dirac framework to derive the boundary physics of the respective topological phase.}
\end{enumerate}

\section{Crystalline topology in the Su-Schrieffer-Heeger model} \label{sec: SSH_overarching}
In this section, we introduce the inversion-symmetric Su-Schrieffer-Heeger Hamiltonian as the elementary topological insulator in one spatial dimension, and use its Dirac approximation (derived in Sec.~\ref{sec: dirac_ham_ssh_derivation}) to study its anomalous boundary physics. 

While previous tutorials have discussed the SSH model as a topological phase protected by chiral symmetry~\cite{AsbothIntro,BernevigNeupertLectures,HasanKaneColloq}, here we instead focus its role as the simplest topological crystalline insulator, protected by inversion symmetry alone~\cite{Alexandradinata14}.

\begin{figure}[t]
\centering
\includegraphics[width=0.55\textwidth,page=3]{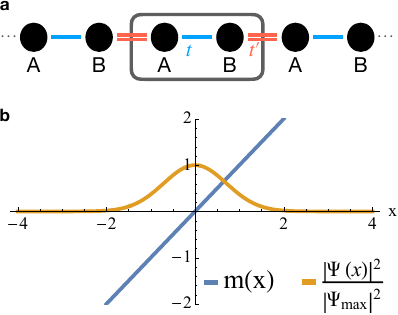}
\caption{Domain-wall bound states in the Su-Schrieffer-Heeger model. (a)~The unit cell contains two atomic sites, A and B, with alternating hopping amplitudes $t \neq t'$ for intra- and inter-unit cell hoppings. (b)~A mass domain wall in the Dirac approximation of the SSH model [Eq.~\eqref{eq: SSHDiracHamiltonian}], modeled by a linear mass profile $m(x) = x$ and shown in blue, gives rise to an exponentially-localized domain wall bound state, shown in orange.
}
\label{fig: ssh_model_lattice_dw}
\end{figure}

\subsection{Motivation of the Hamiltonian} \label{subsec: polyacetylene}
Consider polyacetylene, a polymer chain of $\mathrm{C_2H_2}$ molecules. To minimize potential energy, the structure exhibits an alternation of weak and strong bonds between the two carbon atoms of the unit cell, A and B, which effectively each provide one atomic site that electrons can occupy [see Fig.~\ref{fig: ssh_model_lattice_dw}~(a)]. At half filling, there is therefore one electron per unit cell. The Hamiltonian contains two terms: the weak, intra-unit cell hopping $t$ that connects $A$ with $B$, and the strong, inter-unit cell hopping $t'$ that connects $B$ with the $A$ atom in the next unit cell. Inversion symmetry maps one carbon atom to another, and should therefore again be represented by $\sigma_x$. Equation~(\ref{eq: 1DgeneralBloch}) becomes
\begin{equation} \label{eq: SSH_tight_binding}
\begin{aligned}
    \mathcal{H}_{\text{SSH}}(k) &= \begin{pmatrix} 0 & t \\ t & 0 \end{pmatrix} + \begin{pmatrix} 0 & 0 \\ t' & 0 \end{pmatrix} e^{\mathrm{i} k} + \begin{pmatrix} 0 & t' \\ 0 & 0 \end{pmatrix} e^{-\mathrm{i} k} \\&= (t + t' \cos k) \sigma_x + t' \sin k \sigma_y \\&= (\Delta + \cos k) \sigma_x + \sin k \sigma_y,
\end{aligned}
\end{equation}
where in the last line we identified $\Delta=t$, and set $t' = 1$ in order to recover Eq.~(\ref{eq: SSHmodel}). We have seen in Sec.~\ref{subsec: inversion_topology_general} that the spectrum of $\mathcal{H}_{\text{SSH}}(k)$ is fully gapped when $|\Delta| \neq 1$, or equivalently when $|t| \neq |t'|$ (when there is a mismatch in bond strength). Furthermore, we identified the regime $|\Delta|<1$ as topological, and derived the effective Dirac Hamiltonian, Eq.~(\ref{eq: SSHDiracHamiltonian}), that captures the sign flip of inversion eigenvalues at $k = 0$ as $\Delta$ is tuned. For convenience, we reproduce the Dirac Hamiltonian here:
\begin{equation} \label{eq: ssh_dirac_approx}
    \mathcal{H}_{\text{Dirac}}(k) = m \sigma_x + k \sigma_y,
\end{equation}
Keeping in mind $m = \Delta + 1$ and that we fixed $\Delta < -1$ as the trivial reference phase, we can identify the parameter range $m<0$ as trivial and $m>0$ as topological in the Dirac approximation.

\subsection{Derivation of the domain wall bound state} \label{subsec: ssh_edgestate_derivation}
The Dirac model $\mathcal{H}_{\text{Dirac}}(k)$, just like the original tight-binding Hamiltonian $\mathcal{H}_{\text{SSH}}(k)$, is gapped at all values of $k$, and therefore bulk-insulating. To study the effect of boundaries in the topological phase, we need to model an interface between the trivial and topological regimes. This is most easily achieved by assigning a position-dependence to the mass $m \rightarrow m(x) = m x$, with the convention $m > 0$, so that $m(x)$ is negative (trivial) in the left half of the system, and positive (topological) in the right half. The particular form of the real space dependence beyond such a sign flip is arbitrary -- we might as well have chosen $m(x) = m \mathrm{sign}(x)$, for instance. The reason is that we are only interested in topological properties that survive continuous deformations on either side of the gap closing point at $x=0$. The only notable effect of our particular choice $m(x) = m x$ is that it is differentiable and so gives smoother results than, for instance, $m(x) = m \mathrm{sign}(x)$. 

We conclude that the following real-space Hamiltonian effectively describes the interface between two SSH chains, one trivial and one topological:
\begin{equation} \label{eq: interfaceDiracHamiltonianSSH}
    \mathcal{H}_{\text{bdry}} = \sigma_x m x - \mathrm{i} \sigma_y \frac{\partial}{\partial x},
\end{equation}
where we expressed $k = -\mathrm{i} \partial/\partial x$ as an operator acting on real space, because its eigenvalue $k$ is not anymore a good quantum number [any non-constant $m(x)$ breaks translation symmetry].

Recall the heuristic argument from Sec.~\ref{sec: introduction} that posited that the boundary between a topological and a trivial phase hosts a phase transition, characterized by a gapless energy spectrum. [This argument can be made mathematically rigorous~\cite{GurarieGreens}, here we will find it more illuminating to derive it in the Dirac approximation context.] Guided by this intuition, we will attempt to solve for zero-energy states at the interface, which are the two-component wavefunctions $\ket{\Psi (x)} = [\psi(x),\phi(x)]^\mathrm{T}$ that satisfy the equation
\begin{equation} \label{eq: realspaceSSHzeromode}
    \mathcal{H}_{\text{bdry}} \ket{\Psi (x)} = 0 \iff \frac{\partial}{\partial x} \ket{\Psi (x)} = - \sigma_z m x \ket{\Psi (x)}.
\end{equation}
There is only one normalizable solution to this equation: 
\begin{equation} \label{eq: SSHzeromodesolution}
    \ket{\Psi (x)} = \frac{1}{\mathcal{N}} e^{- \frac{1}{2} m x^2} \begin{pmatrix} 1 \\ 0\end{pmatrix},
\end{equation}
where $\mathcal{N} = (\pi/m)^{\frac{1}{4}}$ is a normalization factor. [The solution that multiplies the vector $(0,1)^\mathrm{T}$ comes with a plus sign in the exponential, and is therefore not normalizable.] In conclusion, we find a single zero-energy state that is exponentially localized to the boundary of the topological phase. Its wavefunction decay is shown in Fig.~\ref{fig: ssh_model_lattice_dw}~(b).

\subsection{Protection by sublattice symmetry: end zeromodes}
What remains is to interpret the zero-energy end state from the point of view of topology. At this point, it is illuminating to first recount the historical point of view, according to which the SSH model realizes a topological phase protected by a local rather than a crystalline symmetry~\cite{BernevigNeupertLectures}. [These conceptually distinct interpretations of topology are both consistent with the Hamiltonian $\mathcal{H}_{\text{SSH}}(k)$, that is, there is more than one symmetry that prevents adiabatic transformations of the SSH ground state to the trivial reference state.] This local symmetry is a sublattice symmetry $\mathcal{C} = \sigma_z$ (also known as chiral symmetry) of $\mathcal{H}_{\text{SSH}}(k)$, and it is realized via 
\begin{equation} \label{eq: sublatticeSymmetryConstraint}
\mathcal{C} \mathcal{H}_{\text{SSH}}(k) \mathcal{C}^\dagger = - \mathcal{H}_{\text{SSH}}(k).
\end{equation}
This relation enforces that the Hamiltonian only couples the two inequivalent sites of the unit cell (the two sublattices) to each other, but not to themselves (as would be the case for local potentials or longer-range hoppings). [Somewhat confusingly, on the level of the Bloch Hamiltonian, sublattice symmetry implies anti-commutation rather than commutation with $\mathcal{C}$. It is nevertheless a valid symmetry of the physical system, because it commutes with the many-body Hamiltonian, Eq.~(\ref{eq: manybodyhamiltonian}) -- the two minus signs, one deriving from the anti-commutation, and another from the fermionic exchange statistics, cancel.]

Sublattice symmetry is preserved by the Dirac approximation and the interface geometry in Eq.~(\ref{eq: interfaceDiracHamiltonianSSH}). Crucially, it implies that for every eigenstates of $\mathcal{H}_{\text{bdry}}$ at energy $E$, with $|E| > 0$, there exists a partner eigenstate at energy $-E$. Only a state at $E=0$ may be unpaired. But this constraint ensures that the single zero-energy mode that we derived in Eq.~(\ref{eq: SSHzeromodesolution}) cannot be moved away from $E=0$ by any local perturbation: a single eigenstate cannot continuously split into two. The edge of a nontrivial SSH insulator therefore hosts a zero-energy bound state that is topologically protected by the bulk gap and sublattice symmetry.\footnote{In fact, sublattice symmetry stabilizes an arbitrary number of end states, and $\mathcal{C}$-symmetric Hamiltonians that host more than one end-localized zeromode exist -- for instance, we may just double all degrees of freedom and hopping elements per unit cell to arrive at a model with two protected end states. We do not consider such models in our tutorial because we want to focus on the more physically relevant protection by inversion symmetry, with which only a single zeromode is stable.}

\subsection{Protection by inversion symmetry: filling anomaly} \label{sec: filling_anomaly}
Let us now return to crystalline symmetries. Whereas it is difficult to guarantee sublattice symmetry in realistic systems, inversion symmetry is a property of many naturally arising crystal structures. We understood in Sec.~\ref{sec: diracapproxintro} that the SSH model also realizes an inversion symmetry-protected topological insulator, and derived the presence of an end state in Eq.~(\ref{eq: SSHzeromodesolution}). It is then natural to ask what properties of the end state are universal and guaranteed by inversion symmetry alone, that is, not specific to the particular choice of Hamiltonian $\mathcal{H}_{\text{SSH}}(k)$. Clearly, absent sublattice symmetry there is no preferred notion of ``zero energy"\footnote{Recall that sublattice symmetry maps an energy $E$ to $-E$, so that the point $E=0$ is special.}, so that lying at zero-energy cannot be a universal property of the domain wall bound state. 

At this point, it is important to note that in order to connect bulk crystalline topology to boundary states, we must study a boundary termination that preserves inversion symmetry. Our choice of the mass profile $m(x)$ from before explicitly broke inversion symmetry -- we only modelled one (the left) end of the topological phase. In order to restore inversion symmetry, we must also take into account the second zero-energy state that is located at the opposite (the right) end of the sample. The inversion-symmetric and finite system will then have two zero-energy states, but as explained above, nothing prevents the addition of a local edge potential to $\mathcal{H}_{\text{SSH}}(k)$ that shifts them to finite energies (at least as long as the local potential is the same on the left and right end of the sample, such a manipulation respects inversion symmetry). The property that remains invariant is therefore not the energy of the end states, but their filling: imagine that we gradually tune a finite system from trivial to topological in an $\mathcal{I}$-symmetric manner. This process can for instance be modeled via the Hamiltonian
\begin{equation} \label{eq: invsymmdomainwallgeometry}
\mathcal{H}_{\text{bdry,}\mathcal{I}}(T) =  \sigma_x m(x,T) - \mathrm{i} \sigma_y \frac{\partial}{\partial x},
\end{equation}
where $T \in [0,1]$, 
\begin{equation}
m(x,0) = -m (m > 0), \quad m(x,1) = - (|x|-L/2) m, 
\end{equation}
and $m(x,T)$ interpolates continuously between $m(x,0)$ and $m(x,1)$ in an inversion-symmetric fashion: $m(x,T) = m(-x,T)$. $\mathcal{H}_{\text{bdry,}\mathcal{I}}(0)$ then describes a fully trivial system without boundaries, while $\mathcal{H}_{\text{bdry,}\mathcal{I}}(1)$ describes a topological system of length $L$ that is embedded in a trivial environment. For large $L$, we may take over the result we obtained for a single edge, Eq.~(\ref{eq: SSHzeromodesolution}), and deduce that $\mathcal{H}_{\text{bdry,}\mathcal{I}}(1)$ hosts two inversion-related end states as discussed above. The only important feature of the interpolating Hamiltonian $\mathcal{H}_{\text{bdry,}\mathcal{I}}(T)$ is that it also respects sublattice symmetry at all values of $T$, ensuring that the two end states that are present at $T=1$ but not at $T=0$ must come in with opposite energies $E_1 (T) = - E_2 (T)$ as $T$ is tuned from $0$ to $1$, with $E_1(1) = E_2(1) = 0$ [recall Eq.~(\ref{eq: sublatticeSymmetryConstraint}) and the surrounding discussion]. Correspondingly, only one state is filled (occupied by an electron in the system's ground state), namely the one deriving from the negative-energy manifold of states.\footnote{To fix the filling, note that the system at $T=0$ is gapped and we are working at zero chemical potential, $\mu = 0$. This implies that the many-body ground state is a Slater determinant of the form of Eq.~(\ref{eq: slaterdetgroundstate}), containing only the negative-energy eigenstates. As $T$ is tuned, it continuously evolves into the ground state at $T=1$, which has two end states at zero energy (as long as the separation $L$ between the two ends is sufficiently large). The filling (particle number in the ground state) is conserved by this interpolation, therefore the zeromode that started out at negative-energy must remain filled at $T=1$, while the zeromode that started out at positive energy must remain empty.} We conclude that the system hosts two end states that together have only one electron's worth of electric charge [recall how the total charge $Q$ is defined in Eq.~(\ref{eq: total_charge}) as the number of occupied electrons]. The only inversion-symmetric way to distribute this charge is to assign half a charge to either end of the system, essentially fractionalizing the electronic charge. 

\begin{figure}[t]
\centering
\includegraphics[width=0.55\textwidth,page=4]{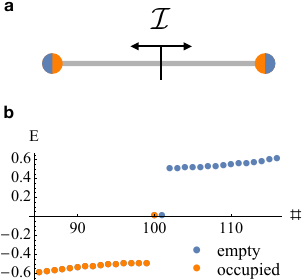}
\caption{Filling anomaly of the inversion-symmetric SSH model with open boundary conditions. (a)~In the topological phase, each end of an inversion-symmetric sample hosts a single bound state worth of half an electronic charge. (b)~Exact diagonalization spectrum of the SSH tight-binding model, Eq.~\eqref{eq: SSH_tight_binding}, with open boundary conditions (L=100). Only the states close to $E=0$ are shown. The presence of two midgap states in the exact model validates the Dirac approximation [discussion around Eq.~\eqref{eq: invsymmdomainwallgeometry}]. Half-filling implies that there are $L$ occupied states, highlighted in orange. Importantly, only one of the two midgap bound states is occupied, giving rise to a filling anomaly.
}
\label{fig: filling_anomaly}
\end{figure}

The tradeoff between the global $U(1)$ symmetry, which naively implies charge quantization for a Hamiltonian of the form of Eq.~(\ref{eq: realspaceHamiltonian}), and the crystalline inversion symmetry, constitutes the so-called \emph{filling anomaly}~\cite{benalcazar2018quantization}. It is the only universal boundary characteristic of inversion-protected topological insulators in one dimension. [See Fig.~\ref{fig: filling_anomaly}~(a)]. The spectrum of Eq.~\eqref{eq: SSH_tight_binding} with open boundary conditions, shown in Fig.~\ref{fig: filling_anomaly}~(b), confirms these results.

\begin{figure*}[t]
\centering
\includegraphics[width=0.98\textwidth,page=5]{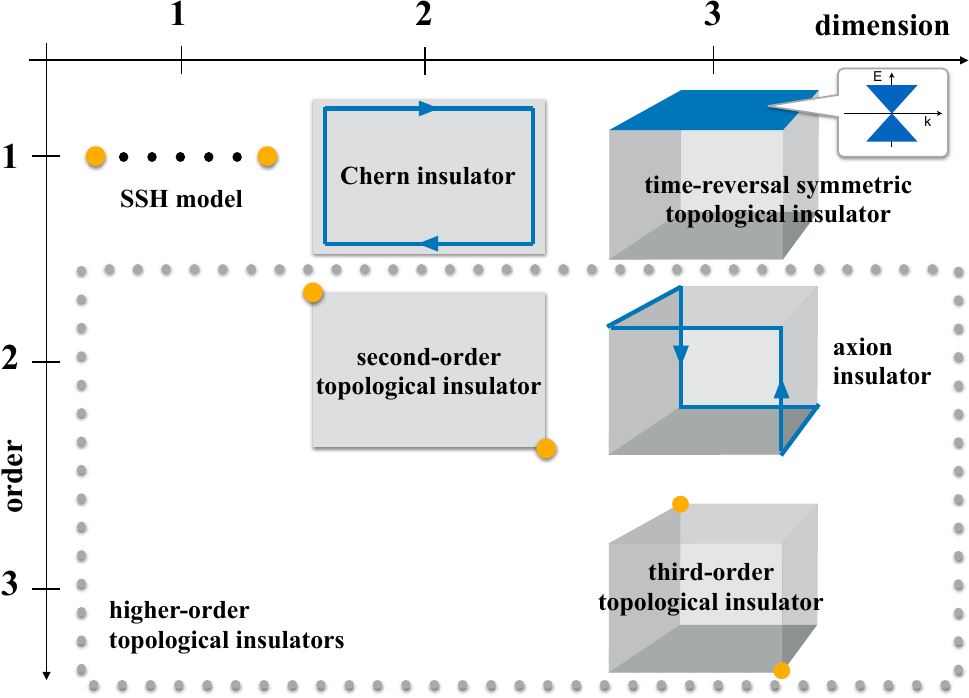}
\caption{Archetypes of topological insulators. The SSH model, as well as the second- and third-order topological insulators, host 0D states on their boundaries (here, the inversion-symmetric case is shown in all three dimensions -- let us note that in 2D and 3D, there are further crystalline symmetries beyond inversion that stabilize higher-order topology). The boundaries of Chern and axion insulators~\cite{Fang:2015aa} host 1D conducting states (1D chiral metals). The surface of the 3D time-reversal symmetric topological insulator features a gapless Dirac cone surface state. Importantly, although the gapless boundary states of $n$-th order phases in $n$ dimensions [and $(n-1)$-th order phases in $n$ dimensions] are of the same type, these phases are genuinely distinct: for instance, a second-order topological insulator in 2D is not adiabatically related to a stack of 1D SSH models.
}
\label{fig: archetypes}
\end{figure*}

While we have used sublattice symmetry to show that the edge states of $\mathcal{H}_{\text{SSH}}(k)$ are (filling) anomalous, it should be emphasized that sublattice symmetry is \emph{not} required to stabilize the anomaly: we used sublattice symmetry simply to shortcut an otherwise lengthy calculation of the domain wall charge. It can be broken without destroying either the crystalline bulk topology or the edge state filling anomaly, which are both protected by inversion symmetry and $U(1)$ charge conservation alone. Note also that the filling anomaly, even though it derives from the zero-dimensional edge states, crucially relies on the topologically nontrivial one-dimensional bulk: as we have shown in Sec.~\ref{sec: tb_basics}, there is no well-defined (regularizable) Hamiltonian that preserves $U(1)$ symmetry and has a ground state with fractional charge.

\section{Second-order topological insulators in two dimensions}
We next leverage what we have learned from the SSH model to understand the simplest possible example of a higher-order topological insulator: the inversion-symmetric case in two spatial dimensions. Again, the Dirac approximation provides the most direct route to a crystalline bulk-boundary correspondence and filling anomaly~\cite{BenHOFAs}.

Historically, higher-order phases were discovered much later~\cite{Benalcazar16,Song17,Langbehn17,Benalcazar17,SchindlerHOTI} than their first-order counterparts such as the Chern insulator~\cite{Haldane88} or the time-reversal symmetric topological insulator~\cite{Kane05a}. Ironically however, at least in 2D, the higher-order case is conceptually simpler. This allows us to bypass the discussion of first-order phases, which have been extensively reviewed elsewhere~\cite{fradkin_2013, AsbothIntro, BernevigBook, BernevigNeupertLectures,HasanKaneColloq}.

\subsection{Concept} \label{subsec: soti_2d_concept}
In Sec.~\ref{sec: SSH_overarching}, we have seen how zero-dimensional states can arise at the ends of a one-dimensional insulator that preserves inversion symmetry. This kind of topological bulk-boundary correspondence is called first-order, because the gapless boundaries have a dimension that is one lower than that of the bulk. Similarly, higher-order topological insulators in $D$ spatial dimensions are defined by the presence of gapless states along boundary segments with dimension $D-K$, where $K > 1$ is the order. The simplest example are corner states in two dimensions, which realize a corner (zero-dimensional) filling anomaly similar to what was discussed in Sec.~\ref{sec: filling_anomaly}, while the bulk (2D) and edges (1D) are gapped. The filling anomaly gives rise to fractional corner charge, similar to how the SSH filling anomaly gave rise to fractional end charge. There also exist 2D first-order topological insulators, most notably the Chern insulator (integer quantum Hall effect) and the time-reversal symmetric topological insulator (quantum spin Hall effect). An overview of first- and higher-order topological phases is given in Fig.~\ref{fig: archetypes}.

First-order phases realize insulators (gapped band structures) in the bulk and anomalous metals [gapless band structures that cannot be obtained from well-defined, regularizable $(D-1)$-dimensional Hamiltonians] on the boundary. Second-order phases, on the other hand, realize insulators in the bulk and \emph{anomalous insulators} [gapped band structures that cannot be obtained from well-defined, regularizable $(D-1)$-dimensional Hamiltonians] on the boundary. These anomalous boundaries can then themselves be viewed as first-order topological phases that host their own gapless end states of dimension $D-2$. 

Crucially, we cannot simply view the $(D-1)$-dimensional boundaries as first-order topological insulators that have been glued to the boundary of an otherwise trivial $D$-dimensional system~\cite{LukaReview}: even though they are gapped, they are anomalous and can only be realized in presence of a topologically nontrivial bulk. These qualitative considerations will be made concrete in the following.

\subsection{Bloch and Dirac Hamiltonian}
Unlike in Sec.~\ref{subsec: polyacetylene}, where we made at least some attempt at motivating the SSH Hamiltonian from the microscopic chemistry of polyacetylene, we will from now on study the properties of toy model Hamiltonians without further physical legitimization. The reason is that these capture the essential physics without distracting us too much with the non-universal properties of any given material. Moreover, higher-order topological phases were only discovered recently, and as of yet only very few ``nice" material realizations are known~\cite{Tang_2019,Vergniory_2019,Zhang_2019}. For our purposes, it will therefore be sufficient to study these novel phases of matter on a conceptual level -- it is one of the great luxuries of condensed matter theory that very often initially theoretical concepts eventually do turn out to be realized in nature.

Following up on our general characterization of two-dimensional second-order topological insulators, let us therefore ask the question: how do we construct a two-dimensional tight-binding that hosts corner states? Evidently, we will again need a crystalline symmetry to ensure topological protection: with only local symmetries in place, any corner states could be brought together along the sample boundary and annihilated with each other -- this process would be adiabatic with respect to the bulk gap. We note that the requirement of crystalline symmetries (symmetries that do not leave all spatial positions $x$ invariant) is particular to higher-order phases. First-order phases \emph{can} be protected by crystalline symmetry operations, but do not \emph{have to} -- we have seen this already for the SSH insulator that can be stabilized by either sublattice or inversion symmetry. For simplicity, we will again consider inversion symmetry here, which in 2D maps both coordinates $\bm{x} = (x,y) \rightarrow (-x,-y)$ to their opposite values.

\begin{figure}[t]
\centering
\includegraphics[width=0.65\textwidth,page=6]{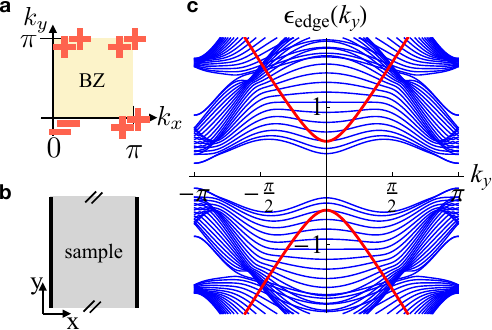}
\caption{Bulk and edge band structure of the 2D higher-order topological insulator modeled by Eq.~\eqref{eq: 2dsoti_tightbindingmodel}. (a)~Bulk inversion eigenvalues in the 2D Brillouin zone. There is a double band inversion at $\mathbf{k} = (0,0)$. (b)~Inversion-symmetric slab geometry with open boundary conditions in the $x$ direction and periodic boundary conditions in the $y$ direction. (c)~Exact diagonalization spectrum of the tight-binding model, Eq.~\eqref{eq: 2dsoti_tightbindingmodel}, in the slab geometry with $L_x = 20$ (shown in blue). The dispersion of the edge states derived from the Dirac approximation, Eq.~\eqref{eq: 2dsoti_edge_hamiltonian}, is shown in red.
}
\label{fig: 2dhoti_bulk_edge}
\end{figure}

In particular, inversion symmetry maps the one-dimensional boundary of a two-dimensional, inversion-symmetric sample to itself. A natural starting point to construct a second-order topological insulator is then to engineer a phase that on its boundary hosts a phase akin to the one-dimensional, inversion-symmetric SSH model -- domain walls in the SSH parameter $m$ [Eq.~(\ref{eq: ssh_dirac_approx})] then bind zero-dimensional ``corner" states. We will show in the following that such a dimensional hierarchy is achieved by the four-band tight-binding Hamiltonian
\begin{equation} \label{eq: 2dsoti_tightbindingmodel}
\begin{aligned}
    \mathcal{H}_{\text{SOTI}}(\bm{k}) =& (\Delta + \cos k_x + \cos k_y) \tau_x \sigma_0 + \sin k_x \tau_z \sigma_x \\& + \sin k_y \tau_z \sigma_y + \delta O,
\end{aligned}
\end{equation}
where $\tau_i$ and $\sigma_i$, $i = 0,x,y,z$ are two sets of Pauli matrices, and we abbreviated the Kronecker product by $\tau_i \otimes \sigma_j \equiv \tau_i \sigma_j$. Inversion symmetry is given by $\mathcal{I} = \tau_x\sigma_0$, and the topological parameter regime is delimited by $|\Delta| < 2$. The small constant $\delta \ll 1$ multiplies a collection of symmetry-allowed perturbation terms $O$ that will be essential in gapping out the one-dimensional boundary.\footnote{In principle, we can add any inversion-symmetric term, $k$-dependent or independent, to the Hamiltonian without changing any of the topological properties (assuming the term is small enough so as not to close the bulk gap). In doing so, we have to tread a fine line: adding no perturbations at all is dangerous because it may give a fine-tuned model that exhibits accidental properties not intrinsic to the topological phase itself, while adding too many perturbing terms makes an analytic treatment burdensome, and thereby hinders an intuitive understanding of the physics. Hence, equipped with the benefit of hindsight, we only consider absolutely essential perturbations in this tutorial. In principle, all symmetry-allowed perturbations will be present in realistic materials, however, they cannot affect topological properties as long as they are so small that they do not close the bulk gap.} We may again define inversion eigenvalues at the four high-symmetry momenta $\bar{\bm{k}} = (0,0), (0,\pi), (\pi,0), (\pi,\pi)$, just as was done in Sec.~\ref{subsec: inversion_topology_general}. Declaring $\Delta < -2$ as the trivial reference phase\footnote{For $\Delta < -2$, all eight occupied inversion eigenvalues (there are two occupied bands and four inversion-symmetric momenta) are equal to $+1$, this configuration is compatible with an atomic limit that is constructed from two occupied $s$ orbitals located at the center of the unit cell.}, the topological phase is then characterized by a sign flip of two occupied inversion eigenvalues at $\bar{\bm{k}} = (0,0)$ [see Fig.~\ref{fig: 2dhoti_bulk_edge}~(a)], prompting us to expand around that momentum in a Dirac approximation:
\begin{equation}
\mathcal{H}_{\text{2D,Dirac}}(\bm{k}) = m \tau_x \sigma_0 + k_x \tau_z \sigma_x + k_y \tau_z \sigma_y + \delta O.
\end{equation}
This Dirac Hamiltonian captures the trivial-to-topological gap-closing transition of $\mathcal{H}_{\text{SOTI}}(\bm{k})$ when $m = \Delta + 2$ is tuned from negative to positive values.

\subsection{Gapped edge states} \label{subsec: gapped_2dsoti_edges}
As for the SSH model, we will first derive the band structure of $\mathcal{H}_{\text{SOTI}}(\bm{k})$ in the presence of a single edge before arguing for the topological signature of a full inversion symmetric boundary. Without loss of generality, we keep periodic boundary conditions in $y$ direction, so that the edge under consideration has its normal along the (negative) $x$ direction. A mass profile $m(\bm{x}) = m x$, $m>0$ then locates the left edge of the topological phase at $x = 0$. Similar to Eq.~(\ref{eq: interfaceDiracHamiltonianSSH}), any boundary-localized zero-energy states are determined by the Hamiltonian
\begin{equation} \label{eq: 2dsoti_edgeham}
    \mathcal{H}_{x}(k_y) = \tau_x \sigma_0 m x - \mathrm{i} \tau_z \sigma_x \frac{\partial}{\partial x} + k_y \tau_z \sigma_y + \delta O.
\end{equation}
There are two normalizable solutions to the zero-energy condition $\mathcal{H}_{x}(0) \ket{\Psi (x)} = 0$ at $k_y = \delta = 0$,
given by
\begin{equation}
    \ket{\Psi_1 (x)} = \frac{1}{\mathcal{N}} e^{- \frac{1}{2} m x^2} \begin{pmatrix} \mathrm{i} \\ 0 \\ 0 \\ 1\end{pmatrix}, \quad \ket{\Psi_2 (x)} = \frac{1}{\mathcal{N}} e^{- \frac{1}{2} m x^2} \begin{pmatrix} 0 \\ \mathrm{i} \\ 1 \\ 0\end{pmatrix},
\end{equation}
where $\mathcal{N} = (4\pi/m)^{\frac{1}{4}}$ is a normalization factor. Note that we chose an ordering convention so that, for instance, the Kronecker product $\tau_x \sigma_0$ is represented as
\begin{equation}
\tau_x \sigma_0 = \begin{pmatrix} 0 & 0 & 1 & 0 \\ 0 & 0 & 0 & 1 \\ 1 & 0 & 0 & 0 \\ 0 & 1 & 0 & 0 \end{pmatrix}.
\end{equation}

We may now deduce the form of the edge Hamiltonian (the Hamiltonian governing the edge degrees of freedom) by degenerate first-order perturbation theory in $k_y$ and $\delta$. Let us for the moment fix $O = \tau_0 \sigma_x$ -- this choice preserves inversion symmetry. The edge Hamiltonian is given by the matrix elements of $\mathcal{H}_{x}(k_y)$ in the basis $\{\ket{\Psi_1 (x)}, \ket{\Psi_2 (x)}\}$ spanned by the zero-energy edge states:
\begin{equation} \label{eq: 2dsoti_edge_hamiltonian}
\begin{aligned}
\relax[\mathcal{H}_{\text{edge}}(k_y)]_{m n} =& \int_{-\infty}^{\infty} \mathrm{d}x \bra{\Psi_m (x)} \mathcal{H}_{x}(k_y) \ket{\Psi_n (x)} \\=& [\delta \xi_x + k_y \xi_y]_{m n},
\end{aligned}
\end{equation}
where we introduced another set of Pauli matrices $\xi_i$, $i=0,x,y,z$. We thus find that the edge of $\mathcal{H}_{\text{SOTI}}(\bm{k})$ is governed by the SSH Dirac Hamiltonian, Eq.~(\ref{eq: SSHDiracHamiltonian}), with the identification $m = \delta$. In particular, any finite $|\delta| > 0$ implies that the edge is gapped out, and so there is no protected filling anomaly stemming from the one-dimensional edge states. The edges of a second-order topological insulator are therefore gapped and seemingly trivial. While some choices of the perturbation $O$ do not lead to a gap (for instance, $\tau_0 \sigma_z$), the existence of a single symmetry-allowed gapping term is already enough to trivialize the edge from a topological point of view (also, such a term will be generically present in real materials). The Dirac results are confirmed by a full diagonalization of $\mathcal{H}_{\text{SOTI}}(\bm{k})$ in a slab geometry, see Fig.~\ref{fig: 2dhoti_bulk_edge}~(b) and~(c).

\subsection{Gapless corner states}
We have shown that the Hamiltonian $\mathcal{H}_{\text{SOTI}}(\bm{k})$ is gapped in bulk and on edges with normal in $x$-direction. Similarly, one can show that the states along edges with normal in $y$-direction are gapped and described by a Hamiltonian just like Eq.~(\ref{eq: 2dsoti_edge_hamiltonian}). We might thus conclude that the entire system is gapped and featureless. There is, however, one catch: the edge termination we considered is not inversion symmetric. Recall how we first derived the SSH end state for a single edge in Sec.~\ref{subsec: ssh_edgestate_derivation}, before deducing that an inversion-symmetric pair of edges hosts a pair of edge states of which only one is filled. To determine the anomalous boundary physics of second-order topological insulators, we likewise have to consider an inversion-symmetric edge termination, for instance that of a rectangular sample. A rectangle has four edges (call them left and right, top and bottom) that all come with their respective version of Eq.~(\ref{eq: 2dsoti_edgeham}) -- in Sec.~\ref{subsec: gapped_2dsoti_edges}, we have explicitly treated the left edge only. It does, however, not suffice to repeat the calculation for the remaining three edges separately: due to the gauge freedom in defining the edge states (we may multiply either state with a phase factor, or take arbitrary superpositions, to arrive at an equally valid set of edge states), there would be no meaningful way of comparing the resultant edge Hamiltonians. Instead, we must find the bound states of an arbitrary edge, labelled by an angle $\phi$ -- these then continuously evolve with $\phi$ as we interpolate from one edge to another, and thereby provide a consistent choice of gauge. 

\begin{figure*}[t]
\centering
\includegraphics[width=1\textwidth,page=7]{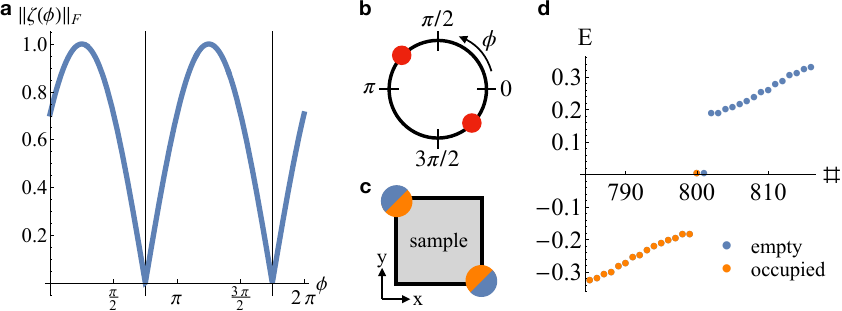}
\caption{Corner states in 2D. (a)~The projected edge mass, Eq.~\eqref{eq: projmass}, has two domain walls at inversion related momenta $\bar{\phi}$ and $\bar{\phi}+\pi$. (b),~(c)~The domain wall bound states at $\bar{\phi}$ and $\bar{\phi}+\pi$ become corner states in a rectangular geometry. (d)~Exact diagonalization spectrum of Eq.~\eqref{eq: 2dsoti_tightbindingmodel}, with open boundary conditions in two directions ($L_x = L_y = 20$). Only the states close to $E=0$ are shown. The half-filled corner midgap states give rise to an inversion symmetry-protected filling anomaly [see discussion around Eq.~\eqref{eq: alternativeO}].
}
\label{fig: cornerstates}
\end{figure*}

For an arbitrary edge that has its normal at an angle $\phi$ to the $x$ axis, it makes sense to decompose the vector of crystal momenta as
\begin{equation} \label{eq: arbitrary_edge_momentum_decomposition}
\bm{k} = k_\perp \begin{pmatrix} \cos \phi \\ \sin \phi \end{pmatrix} + k_\parallel \begin{pmatrix} -\sin \phi \\ \cos \phi \end{pmatrix},
\end{equation}
where $k_\perp$ is the momentum along the edge normal, and $k_\parallel$ the momentum along the edge itself (note that we are orienting $k_\parallel$ at a positive right angle from $k_\perp$). The Dirac Hamiltonian then reads
\begin{equation} \label{eq: DiracHam_generalEdge}
\begin{aligned}
\mathcal{H}^\phi_{\text{2D,Dirac}}&(k_\perp,k_\parallel) = \\&m \tau_x \sigma_0 + k_\perp \tau_z \sigma_1 (\phi) + k_\parallel \tau_z \sigma_2 (\phi) + \delta O,
\end{aligned}
\end{equation}
where $\sigma_1 (\phi) = \sigma_x \cos \phi + \sigma_y \sin \phi$ and $\sigma_2 (\phi) = \sigma_y \cos \phi - \sigma_x \sin \phi$. After decomposing the position coordinate $\bm{x}$ into $x_\perp$ and $x_\parallel$ similar to Eq.~(\ref{eq: arbitrary_edge_momentum_decomposition}), we again implement the edge via a domain wall mass profile $m(x_\perp) = m x_\perp$, $m > 0$, so that the topological phase lies at positive values of $x_\perp$ (the edge normal points into the topological region). This setup retains $k_\parallel$ as a good momentum quantum number, while we replace $k_\perp$ by $-\mathrm{i} \partial/\partial x_\perp$:
\begin{equation}
    \mathcal{H}^\phi_{x_\perp}(k_\parallel) = \tau_x \sigma_0 m x_\perp - \mathrm{i} \tau_z \sigma_1 (\phi) \frac{\partial}{\partial x_\perp} + k_\parallel \tau_z \sigma_2 (\phi) + \delta O.
\end{equation}
Edge states at $k_\parallel = \delta = 0$ then satisfy 
\begin{equation}
[\tau_x \sigma_0 m x_\perp - \mathrm{i} \tau_z \sigma_1 (\phi) \frac{\partial}{\partial x_\perp} ] \ket{\Psi^\phi_n (x_\perp)} = 0.
\end{equation}
Noting that $[\sigma_1 (\phi)] ^2 = \mathbb{1}$, the eigenvalues of $\sigma_1 (\phi)$ are constant and given by $-1$ and $+1$. The equation can then be solved to give
\begin{equation}
\ket{\Psi^\phi_1 (x_\perp)} = \frac{1}{\mathcal{N}} e^{-\frac{1}{2} m x_\perp^2} \begin{pmatrix} e^{-\mathrm{i} \phi} \\ 0 \\ 0 \\ \mathrm{i} \end{pmatrix}, \quad
\ket{\Psi^\phi_2 (x_\perp)} = \frac{1}{\mathcal{N}} e^{-\frac{1}{2} m x_\perp^2} \begin{pmatrix} 0 \\ e^{+\mathrm{i} \phi} \\ \mathrm{i} \\ 0 \end{pmatrix},
\end{equation}
where $\mathcal{N} = (4\pi/m)^{\frac{1}{4}}$ is a normalization factor. The dispersion of these states with edge momentum $k_\parallel$, and their gapping, can again be derived in degenerate first-order perturbation theory by evaluating the matrix elements
\begin{equation} \label{eq: generalphi_edgeham}
\begin{aligned}
\relax[\mathcal{H}^\phi_{\text{edge}}(k_\parallel)]_{m n} =& \int_{-\infty}^{\infty} \mathrm{d}x_\perp \bra{\Psi^\phi_m (x_\perp)} \mathcal{H}^\phi_{x_\perp}(k_\parallel) \ket{\Psi^\phi_n (x_\perp)} \\=& [k_\parallel (\xi_x \sin \phi + \xi_y \cos \phi) + \delta \zeta(\phi)]_{m n},
\end{aligned}
\end{equation}
where 
\begin{equation} 
[\zeta(\phi)]_{m n} = \int_{-\infty}^{\infty} \mathrm{d}x_\perp \bra{\Psi^\phi_m (x_\perp)} O \ket{\Psi^\phi_n (x_\perp)}. 
\end{equation}
Generalizing from Sec.~\ref{subsec: gapped_2dsoti_edges}, we now make the choice
\begin{equation} \label{eq: oconventionfillinganomaly}
O = \tau_0 \sigma_x + \tau_0 \sigma_y,
\end{equation}
so that all edges with normals along the $x$- and $y$-directions are gapped out. We then find
\begin{equation} \label{eq: projmass}
\zeta(\phi) = (\cos \phi + \sin \phi) (\xi_x \cos \phi - \xi_y \sin \phi),
\end{equation}
which anti-commutes with the kinetic term in Eq.~\eqref{eq: generalphi_edgeham} and so forms a mass for the edge states. Importantly however, $\zeta(\phi)$ has two zeroes where all of its entries switch sign. [See Fig.~\ref{fig: cornerstates}~(a).] We conclude that, in a rectangular geometry, the edge states on the left and right edges, as well as on the top and bottom edges, come with opposite signs of their Dirac mass. Viewing the full rectangular boundary as a 1D system governed by a SSH type Hamiltonian [Eq.~\eqref{eq: generalphi_edgeham}], it then hosts Dirac mass domain walls at two oppositely related corners [see Fig.~\ref{fig: cornerstates}~(b),~(c)].\footnote{The presence of two inversion-related zero-dimensional bound states is a general property of the bulk Hamiltonian, and not specific to a rectangular geometry. For instance, we could also have considered a circular geometry with a sufficiently large radius of curvature: as long as a disk-shaped sample is large enough so that we can approximate the local electronic band structure at each point of its boundary by a Dirac equation of the type in Eq.~(\ref{eq: DiracHam_generalEdge}), the two domain walls in $\zeta(\phi)$ give rise to two filling-anomalous bound states on the circle. The failure of our Dirac approximation at small radii has a very physical interpretation: for finite-sized circles, the angular momentum quantum number takes on quantized values, so that in the exact solution of the Dirac equation on a circle there is a finite energy gap in the edge state dispersion, Eq.~(\ref{eq: generalphi_edgeham}), that scales inversely proportional to the radius.} The situation is therefore just that of Eq.~(\ref{eq: invsymmdomainwallgeometry}) at $t=1$, where we considered a SSH model with two inversion-related domain walls, and showed that these host end states with a filling anomaly. Equivalently, the corners of $\mathcal{H}_{\text{SOTI}}(\bm{k})$ bind gapless states that lead to a filling anomaly of second-order: while the infinite slab geometry of Fig.~\eqref{fig: 2dhoti_bulk_edge}~(b) is gapped, a filling anomaly arises when introducing open boundary conditions in two directions. 

Just like the for the SSH model, the filling anomaly is guaranteed by the topologically nontrivial bulk gap and inversion symmetry alone, and neither relies on the particular structure of $\mathcal{H}_{\text{SOTI}}(\bm{k})$ (including the choice of $O$), nor on the rectangular sample geometry we chose to derive it. In particular, while the presence of half-filled midgap states does imply a filling anomaly, the presence of a filling anomaly does not necessitate midgap states (this would only be the case if sublattice symmetry was preserved in addition to inversion symmetry). For instance, consider the choice 
\begin{equation} \label{eq: alternativeO}
O = \tau_0 \sigma_x + \tau_0 \sigma_y + \tau_x \sigma_x + \tau_x \sigma_y,
\end{equation}
which yields a filling anomaly but no midgap states [the corresponding $\zeta(\phi)$ does not have zeroes]. Similar to the sublattice symmetry of Eq.~\eqref{eq: invsymmdomainwallgeometry}, the choice of Eq.~\eqref{eq: oconventionfillinganomaly} is therefore merely a convenient way to short-cut the derivation of a filling anomaly [indeed, with this choice, $\mathcal{H}^\phi_{\text{edge}}(k_\parallel)$ preserves the sublattice symmetry $\mathcal{C} = \xi_z$, while it is broken by Eq.~\eqref{eq: alternativeO}].
Our Dirac results are confirmed by a full diagonalization of the tight-binding Hamiltonian, shown in Fig.~\ref{fig: cornerstates}~(d).

\begin{acknowledgments}
I want to thank Titus Neupert for teaching me most of what is covered in this tutorial, Pierre Fromholz for providing constructive comments, and Daoyuan Li and Poliana Penteado for pointing out errors and inconsistencies in the first version. I am supported by a fellowship at the Princeton Center for Theoretical Science.
\end{acknowledgments}

\bibliography{Ref-Lib}

%apsrmp4-2.bst 2018-12-27 (MD) hand-edited version of apsrmp4-1.bst
%Control: author (3) reversed first dotless
\begin{thebibliography}{32}%
\makeatletter
\providecommand \@ifxundefined [1]{%
 \@ifx{#1\undefined}
}%
\providecommand \@ifnum [1]{%
 \ifnum #1\expandafter \@firstoftwo
 \else \expandafter \@secondoftwo
 \fi
}%
\providecommand \@ifx [1]{%
 \ifx #1\expandafter \@firstoftwo
 \else \expandafter \@secondoftwo
 \fi
}%
\providecommand \natexlab [1]{#1}%
\providecommand \enquote  [1]{``#1''}%
\providecommand \bibnamefont  [1]{#1}%
\providecommand \bibfnamefont [1]{#1}%
\providecommand \citenamefont [1]{#1}%
\providecommand \href@noop [0]{\@secondoftwo}%
\providecommand \href [0]{\begingroup \@sanitize@url \@href}%
\providecommand \@href[1]{\@@startlink{#1}\@@href}%
\providecommand \@@href[1]{\endgroup#1\@@endlink}%
\providecommand \@sanitize@url [0]{\catcode `\\12\catcode `\$12\catcode
  `\&12\catcode `\#12\catcode `\^12\catcode `\_12\catcode `\%12\relax}%
\providecommand \@@startlink[1]{}%
\providecommand \@@endlink[0]{}%
\providecommand \url  [0]{\begingroup\@sanitize@url \@url }%
\providecommand \@url [1]{\endgroup\@href {#1}{\urlprefix }}%
\providecommand \urlprefix  [0]{URL }%
\providecommand \Eprint [0]{\href }%
\providecommand \doibase [0]{https://doi.org/}%
\providecommand \selectlanguage [0]{\@gobble}%
\providecommand \bibinfo  [0]{\@secondoftwo}%
\providecommand \bibfield  [0]{\@secondoftwo}%
\providecommand \translation [1]{[#1]}%
\providecommand \BibitemOpen [0]{}%
\providecommand \bibitemStop [0]{}%
\providecommand \bibitemNoStop [0]{.\EOS\space}%
\providecommand \EOS [0]{\spacefactor3000\relax}%
\providecommand \BibitemShut  [1]{\csname bibitem#1\endcsname}%
\let\auto@bib@innerbib\@empty
%</preamble>
\bibitem [{\citenamefont {Alexandradinata}\ \emph {et~al.}(2014)\citenamefont
  {Alexandradinata}, \citenamefont {Dai},\ and\ \citenamefont
  {Bernevig}}]{Alexandradinata14}%
  \BibitemOpen
  \bibfield  {author} {\bibinfo {author} {\bibnamefont {Alexandradinata},
  \bibfnamefont {A}}, \bibinfo {author} {\bibfnamefont {Xi}~\bibnamefont
  {Dai}}, and\ \bibinfo {author} {\bibfnamefont {B.~Andrei}\ \bibnamefont
  {Bernevig}}} (\bibinfo {year} {2014}),\ \bibfield  {title} {\enquote
  {\bibinfo {title} {Wilson-loop characterization of inversion-symmetric
  topological insulators},}\ }\href
  {https://doi.org/10.1103/PhysRevB.89.155114} {\bibfield  {journal} {\bibinfo
  {journal} {Phys. Rev. B}\ }\textbf {\bibinfo {volume} {89}},\ \bibinfo
  {pages} {155114}}\BibitemShut {NoStop}%
\bibitem [{\citenamefont {Anderson}(1972)}]{Anderson393}%
  \BibitemOpen
  \bibfield  {author} {\bibinfo {author} {\bibnamefont {Anderson},
  \bibfnamefont {P~W}}} (\bibinfo {year} {1972}),\ \bibfield  {title} {\enquote
  {\bibinfo {title} {More is different},}\ }\href
  {https://doi.org/10.1126/science.177.4047.393} {\bibfield  {journal}
  {\bibinfo  {journal} {Science}\ }\textbf {\bibinfo {volume} {177}}~(\bibinfo
  {number} {4047}),\ \bibinfo {pages} {393--396}},\ \Eprint
  {https://arxiv.org/abs/https://science.sciencemag.org/content/177/4047/393.full.pdf}
  {https://science.sciencemag.org/content/177/4047/393.full.pdf} \BibitemShut
  {NoStop}%
\bibitem [{\citenamefont {Asb{\'o}th}\ \emph {et~al.}(2016)\citenamefont
  {Asb{\'o}th}, \citenamefont {Oroszl{\'a}ny},\ and\ \citenamefont
  {P{\'a}lyi}}]{AsbothIntro}%
  \BibitemOpen
  \bibfield  {author} {\bibinfo {author} {\bibnamefont {Asb{\'o}th},
  \bibfnamefont {J{\'a}nos~K}}, \bibinfo {author} {\bibfnamefont
  {L{\'a}szl{\'o}}\ \bibnamefont {Oroszl{\'a}ny}}, and\ \bibinfo {author}
  {\bibfnamefont {Andr{\'a}s}\ \bibnamefont {P{\'a}lyi}}} (\bibinfo {year}
  {2016}),\ \bibfield  {title} {\enquote {\bibinfo {title} {A short course on
  topological insulators},}\ }\href@noop {} {\bibfield  {journal} {\bibinfo
  {journal} {Lecture Notes in Physics}\ }\textbf {\bibinfo {volume}
  {919}}}\BibitemShut {NoStop}%
\bibitem [{\citenamefont {Benalcazar}\ \emph
  {et~al.}(2017{\natexlab{a}})\citenamefont {Benalcazar}, \citenamefont
  {Bernevig},\ and\ \citenamefont {Hughes}}]{Benalcazar17}%
  \BibitemOpen
  \bibfield  {author} {\bibinfo {author} {\bibnamefont {Benalcazar},
  \bibfnamefont {Wladimir~A}}, \bibinfo {author} {\bibfnamefont {B.~Andrei}\
  \bibnamefont {Bernevig}}, and\ \bibinfo {author} {\bibfnamefont {Taylor~L.}\
  \bibnamefont {Hughes}}} (\bibinfo {year} {2017}{\natexlab{a}}),\ \bibfield
  {title} {\enquote {\bibinfo {title} {Electric multipole moments, topological
  multipole moment pumping, and chiral hinge states in crystalline
  insulators},}\ }\href {https://doi.org/10.1103/PhysRevB.96.245115} {\bibfield
   {journal} {\bibinfo  {journal} {Phys. Rev. B}\ }\textbf {\bibinfo {volume}
  {96}},\ \bibinfo {pages} {245115}}\BibitemShut {NoStop}%
\bibitem [{\citenamefont {Benalcazar}\ \emph
  {et~al.}(2017{\natexlab{b}})\citenamefont {Benalcazar}, \citenamefont
  {Bernevig},\ and\ \citenamefont {Hughes}}]{Benalcazar16}%
  \BibitemOpen
  \bibfield  {author} {\bibinfo {author} {\bibnamefont {Benalcazar},
  \bibfnamefont {Wladimir~A}}, \bibinfo {author} {\bibfnamefont {B.~Andrei}\
  \bibnamefont {Bernevig}}, and\ \bibinfo {author} {\bibfnamefont {Taylor~L.}\
  \bibnamefont {Hughes}}} (\bibinfo {year} {2017}{\natexlab{b}}),\ \bibfield
  {title} {\enquote {\bibinfo {title} {Quantized electric multipole
  insulators},}\ }\href {https://doi.org/10.1126/science.aah6442} {\bibfield
  {journal} {\bibinfo  {journal} {Science}\ }\textbf {\bibinfo {volume}
  {357}}~(\bibinfo {number} {6346}),\ \bibinfo {pages} {61--66}}\BibitemShut
  {NoStop}%
\bibitem [{\citenamefont {Benalcazar}\ \emph {et~al.}(2019)\citenamefont
  {Benalcazar}, \citenamefont {Li},\ and\ \citenamefont
  {Hughes}}]{benalcazar2018quantization}%
  \BibitemOpen
  \bibfield  {author} {\bibinfo {author} {\bibnamefont {Benalcazar},
  \bibfnamefont {Wladimir~A}}, \bibinfo {author} {\bibfnamefont {Tianhe}\
  \bibnamefont {Li}}, and\ \bibinfo {author} {\bibfnamefont {Taylor~L.}\
  \bibnamefont {Hughes}}} (\bibinfo {year} {2019}),\ \bibfield  {title}
  {\enquote {\bibinfo {title} {Quantization of fractional corner charge in
  ${C}_{n}$-symmetric higher-order topological crystalline insulators},}\
  }\href {https://doi.org/10.1103/PhysRevB.99.245151} {\bibfield  {journal}
  {\bibinfo  {journal} {Phys. Rev. B}\ }\textbf {\bibinfo {volume} {99}},\
  \bibinfo {pages} {245151}}\BibitemShut {NoStop}%
\bibitem [{\citenamefont {Bernevig}\ and\ \citenamefont
  {Neupert}(2017)}]{BernevigNeupertLectures}%
  \BibitemOpen
  \bibfield  {author} {\bibinfo {author} {\bibnamefont {Bernevig},
  \bibfnamefont {Andrei}}, and\ \bibinfo {author} {\bibfnamefont {Titus}\
  \bibnamefont {Neupert}}} (\bibinfo {year} {2017}),\ \bibfield  {title}
  {\enquote {\bibinfo {title} {Topological superconductors and category
  theory},}\ }\href@noop {} {\bibinfo  {journal} {Lecture Notes of the Les
  Houches Summer School: Topological Aspects of Condensed Matter Physics}\ ,\
  \bibinfo {pages} {63--121}}\BibitemShut {NoStop}%
\bibitem [{\citenamefont {Bernevig}\ and\ \citenamefont
  {Hughes}(2013)}]{BernevigBook}%
  \BibitemOpen
\bibfield  {journal} {  }\bibfield  {author} {\bibinfo {author} {\bibnamefont
  {Bernevig}, \bibfnamefont {B~Andrei}}, and\ \bibinfo {author} {\bibfnamefont
  {Taylor~L}\ \bibnamefont {Hughes}}} (\bibinfo {year} {2013}),\ \href@noop {}
  {\emph {\bibinfo {title} {Topological insulators and topological
  superconductors}}}\ (\bibinfo  {publisher} {Princeton University
  Press})\BibitemShut {NoStop}%
\bibitem [{\citenamefont {Bradlyn}\ \emph {et~al.}(2017)\citenamefont
  {Bradlyn}, \citenamefont {Elcoro}, \citenamefont {Cano}, \citenamefont
  {Vergniory}, \citenamefont {Wang}, \citenamefont {Felser}, \citenamefont
  {Aroyo},\ and\ \citenamefont {Bernevig}}]{Bradlyn17}%
  \BibitemOpen
  \bibfield  {author} {\bibinfo {author} {\bibnamefont {Bradlyn}, \bibfnamefont
  {Barry}}, \bibinfo {author} {\bibfnamefont {L.}~\bibnamefont {Elcoro}},
  \bibinfo {author} {\bibfnamefont {Jennifer}\ \bibnamefont {Cano}}, \bibinfo
  {author} {\bibfnamefont {M.~G.}\ \bibnamefont {Vergniory}}, \bibinfo {author}
  {\bibfnamefont {Zhijun}\ \bibnamefont {Wang}}, \bibinfo {author}
  {\bibfnamefont {C.}~\bibnamefont {Felser}}, \bibinfo {author} {\bibfnamefont
  {M.~I.}\ \bibnamefont {Aroyo}}, and\ \bibinfo {author} {\bibfnamefont
  {B.~Andrei}\ \bibnamefont {Bernevig}}} (\bibinfo {year} {2017}),\ \bibfield
  {title} {\enquote {\bibinfo {title} {Topological quantum chemistry},}\ }\href
  {http://dx.doi.org/10.1038/nature23268} {\bibfield  {journal} {\bibinfo
  {journal} {Nature}\ }\textbf {\bibinfo {volume} {547}}~(\bibinfo {number}
  {7663}),\ \bibinfo {pages} {298--305}}\BibitemShut {NoStop}%
\bibitem [{\citenamefont {Essin}\ and\ \citenamefont
  {Gurarie}(2011)}]{GurarieGreens}%
  \BibitemOpen
  \bibfield  {author} {\bibinfo {author} {\bibnamefont {Essin}, \bibfnamefont
  {Andrew~M}}, and\ \bibinfo {author} {\bibfnamefont {Victor}\ \bibnamefont
  {Gurarie}}} (\bibinfo {year} {2011}),\ \bibfield  {title} {\enquote {\bibinfo
  {title} {Bulk-boundary correspondence of topological insulators from their
  respective green's functions},}\ }\href
  {https://doi.org/10.1103/PhysRevB.84.125132} {\bibfield  {journal} {\bibinfo
  {journal} {Phys. Rev. B}\ }\textbf {\bibinfo {volume} {84}},\ \bibinfo
  {pages} {125132}}\BibitemShut {NoStop}%
\bibitem [{\citenamefont {Fang}\ and\ \citenamefont {Fu}(2015)}]{Fang:2015aa}%
  \BibitemOpen
  \bibfield  {author} {\bibinfo {author} {\bibnamefont {Fang}, \bibfnamefont
  {Chen}}, and\ \bibinfo {author} {\bibfnamefont {Liang}\ \bibnamefont {Fu}}}
  (\bibinfo {year} {2015}),\ \bibfield  {title} {\enquote {\bibinfo {title}
  {New classes of three-dimensional topological crystalline insulators:
  Nonsymmorphic and magnetic},}\ }\href
  {https://doi.org/10.1103/PhysRevB.91.161105} {\bibfield  {journal} {\bibinfo
  {journal} {Physical Review B}\ }\textbf {\bibinfo {volume} {91}}~(\bibinfo
  {number} {16}),\ 10.1103/PhysRevB.91.161105}\BibitemShut {NoStop}%
\bibitem [{\citenamefont {Fradkin}(2013)}]{fradkin_2013}%
  \BibitemOpen
  \bibfield  {author} {\bibinfo {author} {\bibnamefont {Fradkin}, \bibfnamefont
  {Eduardo}}} (\bibinfo {year} {2013}),\ \href
  {https://doi.org/10.1017/CBO9781139015509} {\emph {\bibinfo {title} {Field
  Theories of Condensed Matter Physics}}},\ \bibinfo {edition} {2nd}\ ed.\
  (\bibinfo  {publisher} {Cambridge University Press})\BibitemShut {NoStop}%
\bibitem [{\citenamefont {Fu}(2011)}]{Fu11}%
  \BibitemOpen
  \bibfield  {author} {\bibinfo {author} {\bibnamefont {Fu}, \bibfnamefont
  {Liang}}} (\bibinfo {year} {2011}),\ \bibfield  {title} {\enquote {\bibinfo
  {title} {Topological crystalline insulators},}\ }\href
  {https://doi.org/10.1103/PhysRevLett.106.106802} {\bibfield  {journal}
  {\bibinfo  {journal} {Phys. Rev. Lett.}\ }\textbf {\bibinfo {volume} {106}},\
  \bibinfo {pages} {106802}}\BibitemShut {NoStop}%
\bibitem [{\citenamefont {Haldane}(1988)}]{Haldane88}%
  \BibitemOpen
  \bibfield  {author} {\bibinfo {author} {\bibnamefont {Haldane}, \bibfnamefont
  {F~D~M}}} (\bibinfo {year} {1988}),\ \bibfield  {title} {\enquote {\bibinfo
  {title} {Model for a quantum {H}all effect without landau levels:
  Condensed-matter realization of the "parity anomaly"},}\ }\href
  {https://doi.org/10.1103/PhysRevLett.61.2015} {\bibfield  {journal} {\bibinfo
   {journal} {Phys. Rev. Lett.}\ }\textbf {\bibinfo {volume} {61}},\ \bibinfo
  {pages} {2015--2018}}\BibitemShut {NoStop}%
\bibitem [{\citenamefont {Hasan}\ and\ \citenamefont
  {Kane}(2010)}]{HasanKaneColloq}%
  \BibitemOpen
  \bibfield  {author} {\bibinfo {author} {\bibnamefont {Hasan}, \bibfnamefont
  {M~Z}}, and\ \bibinfo {author} {\bibfnamefont {C.~L.}\ \bibnamefont {Kane}}}
  (\bibinfo {year} {2010}),\ \bibfield  {title} {\enquote {\bibinfo {title}
  {Colloquium: Topological insulators},}\ }\href
  {https://doi.org/10.1103/RevModPhys.82.3045} {\bibfield  {journal} {\bibinfo
  {journal} {Rev. Mod. Phys.}\ }\textbf {\bibinfo {volume} {82}},\ \bibinfo
  {pages} {3045--3067}}\BibitemShut {NoStop}%
\bibitem [{\citenamefont {Kane}\ and\ \citenamefont {Mele}(2005)}]{Kane05a}%
  \BibitemOpen
  \bibfield  {author} {\bibinfo {author} {\bibnamefont {Kane}, \bibfnamefont
  {C~L}}, and\ \bibinfo {author} {\bibfnamefont {E.~J.}\ \bibnamefont {Mele}}}
  (\bibinfo {year} {2005}),\ \bibfield  {title} {\enquote {\bibinfo {title}
  {{$\mathbb{Z}_2$} topological order and the quantum spin {H}all effect},}\
  }\href {https://doi.org/10.1103/PhysRevLett.95.146802} {\bibfield  {journal}
  {\bibinfo  {journal} {Phys. Rev. Lett.}\ }\textbf {\bibinfo {volume} {95}},\
  \bibinfo {pages} {146802}}\BibitemShut {NoStop}%
\bibitem [{\citenamefont {Langbehn}\ \emph {et~al.}(2017)\citenamefont
  {Langbehn}, \citenamefont {Peng}, \citenamefont {Trifunovic}, \citenamefont
  {von Oppen},\ and\ \citenamefont {Brouwer}}]{Langbehn17}%
  \BibitemOpen
  \bibfield  {author} {\bibinfo {author} {\bibnamefont {Langbehn},
  \bibfnamefont {Josias}}, \bibinfo {author} {\bibfnamefont {Yang}\
  \bibnamefont {Peng}}, \bibinfo {author} {\bibfnamefont {Luka}\ \bibnamefont
  {Trifunovic}}, \bibinfo {author} {\bibfnamefont {Felix}\ \bibnamefont {von
  Oppen}}, and\ \bibinfo {author} {\bibfnamefont {Piet~W.}\ \bibnamefont
  {Brouwer}}} (\bibinfo {year} {2017}),\ \bibfield  {title} {\enquote {\bibinfo
  {title} {Reflection-symmetric second-order topological insulators and
  superconductors},}\ }\href {https://doi.org/10.1103/PhysRevLett.119.246401}
  {\bibfield  {journal} {\bibinfo  {journal} {Phys. Rev. Lett.}\ }\textbf
  {\bibinfo {volume} {119}},\ \bibinfo {pages} {246401}}\BibitemShut {NoStop}%
\bibitem [{\citenamefont {Neupert}\ and\ \citenamefont
  {Schindler}(2018)}]{Neupert2018}%
  \BibitemOpen
  \bibfield  {author} {\bibinfo {author} {\bibnamefont {Neupert}, \bibfnamefont
  {Titus}}, and\ \bibinfo {author} {\bibfnamefont {Frank}\ \bibnamefont
  {Schindler}}} (\bibinfo {year} {2018}),\ \enquote {\bibinfo {title}
  {Topological crystalline insulators},}\ in\ \href
  {https://doi.org/10.1007/978-3-319-76388-0_2} {\emph {\bibinfo {booktitle}
  {Topological Matter: Lectures from the Topological Matter School 2017}}},\
  \bibinfo {editor} {edited by\ \bibinfo {editor} {\bibfnamefont {Dario}\
  \bibnamefont {Bercioux}}, \bibinfo {editor} {\bibfnamefont {J{\'e}r{\^o}me}\
  \bibnamefont {Cayssol}}, \bibinfo {editor} {\bibfnamefont {Maia~G.}\
  \bibnamefont {Vergniory}}, \ and\ \bibinfo {editor} {\bibfnamefont
  {M.}~\bibnamefont {Reyes~Calvo}}}\ (\bibinfo  {publisher} {Springer
  International Publishing},\ \bibinfo {address} {Cham})\ pp.\ \bibinfo {pages}
  {31--61}\BibitemShut {NoStop}%
\bibitem [{\citenamefont {Po}(2020)}]{Po_2020}%
  \BibitemOpen
  \bibfield  {author} {\bibinfo {author} {\bibnamefont {Po}, \bibfnamefont
  {Hoi~Chun}}} (\bibinfo {year} {2020}),\ \bibfield  {title} {\enquote
  {\bibinfo {title} {Symmetry indicators of band topology},}\ }\href
  {https://doi.org/10.1088/1361-648x/ab7adb} {\bibfield  {journal} {\bibinfo
  {journal} {Journal of Physics: Condensed Matter}\ }\textbf {\bibinfo {volume}
  {32}}~(\bibinfo {number} {26}),\ \bibinfo {pages} {263001}}\BibitemShut
  {NoStop}%
\bibitem [{\citenamefont {Po}\ \emph {et~al.}(2018)\citenamefont {Po},
  \citenamefont {Watanabe},\ and\ \citenamefont {Vishwanath}}]{AshvinFragile}%
  \BibitemOpen
  \bibfield  {author} {\bibinfo {author} {\bibnamefont {Po}, \bibfnamefont
  {Hoi~Chun}}, \bibinfo {author} {\bibfnamefont {Haruki}\ \bibnamefont
  {Watanabe}}, and\ \bibinfo {author} {\bibfnamefont {Ashvin}\ \bibnamefont
  {Vishwanath}}} (\bibinfo {year} {2018}),\ \bibfield  {title} {\enquote
  {\bibinfo {title} {Fragile topology and {W}annier obstructions},}\ }\href
  {https://doi.org/10.1103/PhysRevLett.121.126402} {\bibfield  {journal}
  {\bibinfo  {journal} {Phys. Rev. Lett.}\ }\textbf {\bibinfo {volume} {121}},\
  \bibinfo {pages} {126402}}\BibitemShut {NoStop}%
\bibitem [{\citenamefont {Schindler}\ \emph {et~al.}(2018)\citenamefont
  {Schindler}, \citenamefont {Cook}, \citenamefont {Vergniory}, \citenamefont
  {Wang}, \citenamefont {Parkin}, \citenamefont {Bernevig},\ and\ \citenamefont
  {Neupert}}]{SchindlerHOTI}%
  \BibitemOpen
  \bibfield  {author} {\bibinfo {author} {\bibnamefont {Schindler},
  \bibfnamefont {Frank}}, \bibinfo {author} {\bibfnamefont {Ashley~M.}\
  \bibnamefont {Cook}}, \bibinfo {author} {\bibfnamefont {Maia~G.}\
  \bibnamefont {Vergniory}}, \bibinfo {author} {\bibfnamefont {Zhijun}\
  \bibnamefont {Wang}}, \bibinfo {author} {\bibfnamefont {Stuart S.~P.}\
  \bibnamefont {Parkin}}, \bibinfo {author} {\bibfnamefont {B.~Andrei}\
  \bibnamefont {Bernevig}}, and\ \bibinfo {author} {\bibfnamefont {Titus}\
  \bibnamefont {Neupert}}} (\bibinfo {year} {2018}),\ \bibfield  {title}
  {\enquote {\bibinfo {title} {Higher-order topological insulators},}\ }\href
  {https://doi.org/10.1126/sciadv.aat0346} {\bibfield  {journal} {\bibinfo
  {journal} {Science Advances}\ }\textbf {\bibinfo {volume} {4}}~(\bibinfo
  {number} {6}),\ \bibinfo {pages} {eaat0346}}\BibitemShut {NoStop}%
\bibitem [{\citenamefont {Senthil}(2015)}]{senthil15}%
  \BibitemOpen
  \bibfield  {author} {\bibinfo {author} {\bibnamefont {Senthil}, \bibfnamefont
  {T}}} (\bibinfo {year} {2015}),\ \bibfield  {title} {\enquote {\bibinfo
  {title} {Symmetry-protected topological phases of quantum matter},}\ }\href
  {https://doi.org/10.1146/annurev-conmatphys-031214-014740} {\bibfield
  {journal} {\bibinfo  {journal} {Annual Review of Condensed Matter Physics}\
  }\textbf {\bibinfo {volume} {6}}~(\bibinfo {number} {1}),\ \bibinfo {pages}
  {299--324}}\BibitemShut {NoStop}%
\bibitem [{\citenamefont {Song}\ \emph {et~al.}(2017)\citenamefont {Song},
  \citenamefont {Fang},\ and\ \citenamefont {Fang}}]{Song17}%
  \BibitemOpen
  \bibfield  {author} {\bibinfo {author} {\bibnamefont {Song}, \bibfnamefont
  {Zhida}}, \bibinfo {author} {\bibfnamefont {Zhong}\ \bibnamefont {Fang}},
  and\ \bibinfo {author} {\bibfnamefont {Chen}\ \bibnamefont {Fang}}} (\bibinfo
  {year} {2017}),\ \bibfield  {title} {\enquote {\bibinfo {title}
  {$(d\ensuremath{-}2)$-dimensional edge states of rotation symmetry protected
  topological states},}\ }\href
  {https://doi.org/10.1103/PhysRevLett.119.246402} {\bibfield  {journal}
  {\bibinfo  {journal} {Phys. Rev. Lett.}\ }\textbf {\bibinfo {volume} {119}},\
  \bibinfo {pages} {246402}}\BibitemShut {NoStop}%
\bibitem [{\citenamefont {Su}\ \emph {et~al.}(1979)\citenamefont {Su},
  \citenamefont {Schrieffer},\ and\ \citenamefont {Heeger}}]{SSH}%
  \BibitemOpen
  \bibfield  {author} {\bibinfo {author} {\bibnamefont {Su}, \bibfnamefont
  {W~P}}, \bibinfo {author} {\bibfnamefont {J.~R.}\ \bibnamefont {Schrieffer}},
  and\ \bibinfo {author} {\bibfnamefont {A.~J.}\ \bibnamefont {Heeger}}}
  (\bibinfo {year} {1979}),\ \bibfield  {title} {\enquote {\bibinfo {title}
  {Solitons in polyacetylene},}\ }\href
  {https://doi.org/10.1103/PhysRevLett.42.1698} {\bibfield  {journal} {\bibinfo
   {journal} {Phys. Rev. Lett.}\ }\textbf {\bibinfo {volume} {42}},\ \bibinfo
  {pages} {1698--1701}}\BibitemShut {NoStop}%
\bibitem [{\citenamefont {Tang}\ \emph {et~al.}(2019)\citenamefont {Tang},
  \citenamefont {Po}, \citenamefont {Vishwanath},\ and\ \citenamefont
  {Wan}}]{Tang_2019}%
  \BibitemOpen
  \bibfield  {author} {\bibinfo {author} {\bibnamefont {Tang}, \bibfnamefont
  {Feng}}, \bibinfo {author} {\bibfnamefont {Hoi~Chun}\ \bibnamefont {Po}},
  \bibinfo {author} {\bibfnamefont {Ashvin}\ \bibnamefont {Vishwanath}}, and\
  \bibinfo {author} {\bibfnamefont {Xiangang}\ \bibnamefont {Wan}}} (\bibinfo
  {year} {2019}),\ \bibfield  {title} {\enquote {\bibinfo {title}
  {Comprehensive search for topological materials using symmetry indicators},}\
  }\href {https://doi.org/10.1038/s41586-019-0937-5} {\bibfield  {journal}
  {\bibinfo  {journal} {Nature}\ }\textbf {\bibinfo {volume} {566}}~(\bibinfo
  {number} {7745}),\ \bibinfo {pages} {486--489}}\BibitemShut {NoStop}%
\bibitem [{\citenamefont {Thouless}\ \emph {et~al.}(1982)\citenamefont
  {Thouless}, \citenamefont {Kohmoto}, \citenamefont {Nightingale},\ and\
  \citenamefont {den Nijs}}]{Thouless82}%
  \BibitemOpen
  \bibfield  {author} {\bibinfo {author} {\bibnamefont {Thouless},
  \bibfnamefont {D~J}}, \bibinfo {author} {\bibfnamefont {M.}~\bibnamefont
  {Kohmoto}}, \bibinfo {author} {\bibfnamefont {M.~P.}\ \bibnamefont
  {Nightingale}}, and\ \bibinfo {author} {\bibfnamefont {M.}~\bibnamefont {den
  Nijs}}} (\bibinfo {year} {1982}),\ \bibfield  {title} {\enquote {\bibinfo
  {title} {Quantized {H}all conductance in a two-dimensional periodic
  potential},}\ }\href {https://doi.org/10.1103/PhysRevLett.49.405} {\bibfield
  {journal} {\bibinfo  {journal} {Phys. Rev. Lett.}\ }\textbf {\bibinfo
  {volume} {49}},\ \bibinfo {pages} {405--408}}\BibitemShut {NoStop}%
\bibitem [{\citenamefont {Trifunovic}\ and\ \citenamefont
  {Brouwer}(2020)}]{LukaReview}%
  \BibitemOpen
  \bibfield  {author} {\bibinfo {author} {\bibnamefont {Trifunovic},
  \bibfnamefont {Luka}}, and\ \bibinfo {author} {\bibfnamefont {Piet~W.}\
  \bibnamefont {Brouwer}}} (\bibinfo {year} {2020}),\ \bibfield  {title}
  {\enquote {\bibinfo {title} {Higher-order topological band structures},}\
  }\href {https://doi.org/10.1002/pssb.202000090} {\bibfield  {journal}
  {\bibinfo  {journal} {physica status solidi (b)}\ ,\ \bibinfo {pages}
  {2000090}}}\Eprint
  {https://arxiv.org/abs/https://onlinelibrary.wiley.com/doi/pdf/10.1002/pssb.202000090}
  {https://onlinelibrary.wiley.com/doi/pdf/10.1002/pssb.202000090} \BibitemShut
  {NoStop}%
\bibitem [{\citenamefont {Vergniory}\ \emph {et~al.}(2019)\citenamefont
  {Vergniory}, \citenamefont {Elcoro}, \citenamefont {Felser}, \citenamefont
  {Regnault}, \citenamefont {Bernevig},\ and\ \citenamefont
  {Wang}}]{Vergniory_2019}%
  \BibitemOpen
  \bibfield  {author} {\bibinfo {author} {\bibnamefont {Vergniory},
  \bibfnamefont {M~G}}, \bibinfo {author} {\bibfnamefont {L.}~\bibnamefont
  {Elcoro}}, \bibinfo {author} {\bibfnamefont {Claudia}\ \bibnamefont
  {Felser}}, \bibinfo {author} {\bibfnamefont {Nicolas}\ \bibnamefont
  {Regnault}}, \bibinfo {author} {\bibfnamefont {B.~Andrei}\ \bibnamefont
  {Bernevig}}, and\ \bibinfo {author} {\bibfnamefont {Zhijun}\ \bibnamefont
  {Wang}}} (\bibinfo {year} {2019}),\ \bibfield  {title} {\enquote {\bibinfo
  {title} {A complete catalogue of high-quality topological materials},}\
  }\href {https://doi.org/10.1038/s41586-019-0954-4} {\bibfield  {journal}
  {\bibinfo  {journal} {Nature}\ }\textbf {\bibinfo {volume} {566}}~(\bibinfo
  {number} {7745}),\ \bibinfo {pages} {480--485}}\BibitemShut {NoStop}%
\bibitem [{\citenamefont {Wieder}\ \emph {et~al.}(2020)\citenamefont {Wieder},
  \citenamefont {Wang}, \citenamefont {Cano}, \citenamefont {Dai},
  \citenamefont {Schoop}, \citenamefont {Bradlyn},\ and\ \citenamefont
  {Bernevig}}]{BenHOFAs}%
  \BibitemOpen
  \bibfield  {author} {\bibinfo {author} {\bibnamefont {Wieder}, \bibfnamefont
  {Benjamin~J}}, \bibinfo {author} {\bibfnamefont {Zhijun}\ \bibnamefont
  {Wang}}, \bibinfo {author} {\bibfnamefont {Jennifer}\ \bibnamefont {Cano}},
  \bibinfo {author} {\bibfnamefont {Xi}~\bibnamefont {Dai}}, \bibinfo {author}
  {\bibfnamefont {Leslie~M.}\ \bibnamefont {Schoop}}, \bibinfo {author}
  {\bibfnamefont {Barry}\ \bibnamefont {Bradlyn}}, and\ \bibinfo {author}
  {\bibfnamefont {B.~Andrei}\ \bibnamefont {Bernevig}}} (\bibinfo {year}
  {2020}),\ \bibfield  {title} {\enquote {\bibinfo {title} {Strong and fragile
  topological dirac semimetals with higher-order fermi arcs},}\ }\href
  {https://doi.org/10.1038/s41467-020-14443-5} {\bibfield  {journal} {\bibinfo
  {journal} {Nature Communications}\ }\textbf {\bibinfo {volume}
  {11}}~(\bibinfo {number} {1}),\ \bibinfo {pages} {627}}\BibitemShut {NoStop}%
\bibitem [{\citenamefont {Wilczek}(1982)}]{wilczek1982quantum}%
  \BibitemOpen
  \bibfield  {author} {\bibinfo {author} {\bibnamefont {Wilczek}, \bibfnamefont
  {Frank}}} (\bibinfo {year} {1982}),\ \bibfield  {title} {\enquote {\bibinfo
  {title} {Quantum mechanics of fractional-spin particles},}\ }\href@noop {}
  {\bibfield  {journal} {\bibinfo  {journal} {Physical review letters}\
  }\textbf {\bibinfo {volume} {49}}~(\bibinfo {number} {14}),\ \bibinfo {pages}
  {957}}\BibitemShut {NoStop}%
\bibitem [{\citenamefont {Zee}(2010)}]{zee2010quantum}%
  \BibitemOpen
  \bibfield  {author} {\bibinfo {author} {\bibnamefont {Zee}, \bibfnamefont
  {Anthony}}} (\bibinfo {year} {2010}),\ \href@noop {} {\emph {\bibinfo {title}
  {Quantum field theory in a nutshell}}}\ (\bibinfo  {publisher} {Princeton
  university press})\BibitemShut {NoStop}%
\bibitem [{\citenamefont {Zhang}\ \emph {et~al.}(2019)\citenamefont {Zhang},
  \citenamefont {Jiang}, \citenamefont {Song}, \citenamefont {Huang},
  \citenamefont {He}, \citenamefont {Fang}, \citenamefont {Weng},\ and\
  \citenamefont {Fang}}]{Zhang_2019}%
  \BibitemOpen
  \bibfield  {author} {\bibinfo {author} {\bibnamefont {Zhang}, \bibfnamefont
  {Tiantian}}, \bibinfo {author} {\bibfnamefont {Yi}~\bibnamefont {Jiang}},
  \bibinfo {author} {\bibfnamefont {Zhida}\ \bibnamefont {Song}}, \bibinfo
  {author} {\bibfnamefont {He}~\bibnamefont {Huang}}, \bibinfo {author}
  {\bibfnamefont {Yuqing}\ \bibnamefont {He}}, \bibinfo {author} {\bibfnamefont
  {Zhong}\ \bibnamefont {Fang}}, \bibinfo {author} {\bibfnamefont {Hongming}\
  \bibnamefont {Weng}}, and\ \bibinfo {author} {\bibfnamefont {Chen}\
  \bibnamefont {Fang}}} (\bibinfo {year} {2019}),\ \bibfield  {title} {\enquote
  {\bibinfo {title} {Catalogue of topological electronic materials},}\ }\href
  {https://doi.org/10.1038/s41586-019-0944-6} {\bibfield  {journal} {\bibinfo
  {journal} {Nature}\ }\textbf {\bibinfo {volume} {566}}~(\bibinfo {number}
  {7745}),\ \bibinfo {pages} {475--479}}\BibitemShut {NoStop}%
\end{thebibliography}%

\end{document}